%% file: HESSLLDB.tex
\documentclass[showpacs,preprintnumbers,amsmath,amssymb,12pt]{revtex4}

\usepackage{amsmath, amssymb, mathrsfs, bbm, subfigure}
\usepackage{graphicx}
\usepackage{dcolumn}
\usepackage{bm}
\usepackage{color}

\include{macros}

\begin{document}

\title{High-Energy Symmetry of Bosonic Open String Theory\\ in the Light-like Linear Dilaton Background}


\author{Chuan-Tsung Chan}
\email{ctchan@thu.edu.tw}
\author{Wei-Ming Chen}
\email{tainist@gmail.com}
\affiliation{Department of Physics, Tunghai University, Taiwan, 40704}



\date{\today}

\begin{abstract}
 High-energy limits of fixed-angle tree-level stringy scattering amplitudes in the light-like linear dilaton background are calculated. Treating the time component of the gradient of light-like dilaton field ($V_0$) as a moduli parameter, we show that: (1) there exists a new fixed-point ($V_0/E\rightarrow \infty$) in the moduli space of the bosonic open string theory, where a new high-energy symmetry among scattering amplitudes can be identified, (2) this new symmetry can be interpreted as a deformation of the flat-space high-energy symmetry, as proposed by D. Gross. Hence, our results give a concrete illustration about the relation between high-energy stringy symmetry and the background independent formulation of string theory.
\end{abstract}


\maketitle
\clearpage
\section{Introduction}
\input{I}
\section{Bosonic open string theory in the linear dilaton background}
\subsection{Polyakov action for the bosonic open string theory in the linear dilaton background}
\input{II-A}
\subsection{Covariant spectrum of physical states in the bosonic open string theory in the linear dilaton background}
\input{II-B}
\subsection{Vertex operators and string scattering amplitudes of the bosonic open string theory in the linear dilaton background}
\input{II-C}
\section{Covariant spectrum of the first massive spin-two particle in the linear dilaton background}
In this section, based on the method in \cite{Chan:2005qd}, we solve for the Virasoro constraints of bosonic open string theory in the linear dilaton background, and derive a general decomposition of physical states in terms of zero and positive norm states at the first massive level.
\subsection{Construction of the zero-norm states}
\input{III-A}
\subsection{Spectrum of positive-norm states}
\input{III-B}
\subsection{General decomposition of a physical states at $\al^\prime m^2=1$ level}
\input{III-C}
\section{High-energy limits of the stringy scattering amplitudes in the light-like linear dilaton background}
In this section, we discuss the calculations of high-energy limits of stringy scattering amplitudes \cite{Chan:2009xx} in the light-like linear dilaton background. The goal is to examine the deformation of high-energy stringy symmetry as a continuous function of the moduli parameter, namely, the light-like dilaton gradient $V^\mu$. This section consists of four parts: we first define all relevant kinematic variables in part A, and obtain various three-point and the high-energy limits of four-point functions in part B and C. Finally, we identify the replacement rules for high-energy stringy scattering amplitudes and compare the high-energy stringy symmetry at two fix-points of moduli space in part D.
\subsection{Kinematic setup}
Here we list the relevant kinematic variables for the calculations of stringy scattering amplitudes in the light-like linear dilaton background. For simplicity, we shall restrict all momenta to lie in a two dimensional plane.
\subsubsection{Kinematics for three-point functions in the light-like linear dilaton background}
\input{IV-A-1.tex}
\subsubsection{Kinematics for four-point functions in the light-like linear dilaton background}
\input{IV-A-2.tex}
\subsubsection{Polarization vectors for photon $(\al^\prime m^2=0)$}
\input{IV-A-3.tex}
\subsubsection{Mandelstam variables and momemtum-polarization contractions}
\input{IV-A-4.tex}

\subsection{Stringy symmetry of the three-point functions}
Three-point functions are typically fixed by the on-shell conditions and momentum conservation, hence there is no such a concept of high-energy limits. However, in the linear dilaton background, three-point stringy scattering amplitudes are functions of a free parameter $V_0$. Hence, it is of interest to see how these coupling constants evolve from $V_0=0$ (flat space-time) to $V_0/E\rightarrow\infty$ (strong linear dilaton gradient). Here we first list the results from \cite{Chan:2009xx} which consists of seven sample calculations of tree-level three-point functions. Then we rewrite the stringy scattering amplitudes in terms of helicity basis. Finally, we compare two sets of coupling constants at two fixed-points in the moduli space.
\subsubsection{General results for three-point functions in the light-like linear dilaton background}
\input{IV-B-1}
\subsubsection{Three-point functions in the helicity representation}
\input{IV-B-2}
\subsubsection{Symmetry pattern of three-point functions}
\input{IV-B-3}
\subsection{High-energy stringy symmetry of the four-point functions}
In this section, we discuss the high-energy stringy symmetry of four-point functions. Based on the previous results \cite{Chan:2009xx}, we first extract the fixed-angle high-energy limits ($\al^\prime E^2\rightarrow\infty$, $\phi$ fixed) of stringy scattering amplitudes in the flat space-time background by taking $V_0=0$. Here we provide some new explicit check for our master formula in \cite{Chan:2003ee,Chan:2004yz}, and we shall show that the inter-level stringy symmetry can be realized in the form of replacement rules. Next we identify a new fixed-point in the moduli space ($V_0/E\rightarrow \infty$), and extract the high-energy limits of stringy scattering amplitudes at this fixed-point. Finally, we show how the stringy symmetry is deformed from one fixed-point to another by comparing two sets of replacement rules.
\subsubsection{General results for four-point functions in the light-like linear dilaton background}
\input{IV-C-1}
\clearpage
\subsubsection{High-energy stringy limits of four-point functions in the flat space-time background}
\input{IV-C-2}

\subsubsection{High-energy stringy limits of four-point functions in the light-like linear dilaton background}
\input{IV-C-3}
\subsubsection{High-energy stringy symmetry and replacement rules for four-point functions}
\input{IV-C-4}
\section{Summary and Conclusion}
\input{V}

\begin{acknowledgments}
\input{AKG}
\end{acknowledgments}


\end{document}

%% file: macros.tex
\usepackage{bbm}

\newcommand{\beqs}{\begin{equation*}}
\def\beq{\begin{equation}}

\newcommand{\eeqs}{\end{equation*}}
\def\eeq{\end{equation}}

\newcommand{\beqas}{\begin{eqnarray*}}
\newcommand{\beqa}{\begin{eqnarray}}

\newcommand{\eeqas}{\end{eqnarray*}}
\newcommand{\eeqa}{\end{eqnarray}}

\newcommand{\al}{\alpha}
\newcommand{\be}{\beta}

\newcommand{\de}{\delta}

\newcommand{\ka}{\kappa}

\newcommand{\si}{\sigma}

\newcommand{\Ga}{\Gamma}

\newcommand{\Si}{\Sigma}

\newcommand{\blist}{\begin{itemize}}

\newcommand{\elist}{\end{itemize}}

\providecommand{\href}[2]{#2}

\DeclareFontFamily{OT1}{rsfs}{}
\DeclareFontShape{OT1}{rsfs}{m}{n}{ <-7> rsfs5 <7-10> rsfs7 <10->rsfs10}{} 
\DeclareMathAlphabet{\mycal}{OT1}{rsfs}{m}{n}

\def\cV{{\cal V}}
\def\cL{{\cal L}}

\def\cM{{\cal M}}

\def\cA{{\cal A}}

\def\TrL2{{\rm Tr}_{L^2}}

\def\atbdry{\Big|_{\partial \cM}}
\def\atbdry0{\Big|_{\partial \cM_0}}
\def\atbdry1{\Big|_{\partial \cM_1}}

\newcommand{\ket}[1]{| #1 \rangle}

\newcommand{\braket}[2]{\langle #1 | #2 \rangle}

%% file: I.tex
\par Based on a series of work on the high-energy scatterings \cite{Gross:1987kza,Gross:1987ar,Gross:1989ge} which relies on semi-classical (saddle-point) approximation in the functional integral evaluation of stringy scattering amplitudes, D. Gross conjectured that there exists a high-energy symmetry in string theory \cite{Gross:1988ue}.
This is an infinite dimensional symmetry which treats all stringy excitations as a single multiplet and relates high-energy scattering amplitudes among physically inequivalent degrees of freedom. Specifically, there are infinitely many linear relations among high-energy stringy scattering amplitudes of particles at the same mass level, and one can use inter-level symmetry to obtain any four-point amplitudes from the the four-tachyon Veneziano amplitude \cite{Moore:1993zc,Moore:1993qe,Moore:1994rm}.
\par In a later reinvestigation of this problem \cite{Lee:2003vm,Chan:2003ee, Chan:2004yz,Chan:2004tb,Chan:2005ne,Chan:2005zp,Chan:2005ji,Ho:2006zu}, the authors have clarified some of the important issues related to the high-energy stringy symmetry (HESS):
\begin{enumerate}
  \item[(1)] Linear relations among high-energy stringy scattering amplitudes at fixed mass level can be derived algebraically based on decoupling of the high-energy zero-norm states.
 \item[(2)] Leading behaviors of the high-energy stringy scattering amplitudes can be obtained based on saddle-point approximation and it is crucial to keep subleading $\frac{1}{\al^\prime E^2}$ (polynomial) corrections in order to derive linear proportional constants among scattering amplitudes. Also, the explicit formulae of the four-point high-energy stringy scattering amplitudes, as a function of energy and scattering angle, leads to a symmetry pattern for inter-level scattering amplitudes.
\end{enumerate}
\par It is important to note that the high-energy stringy symmetry we discuss here is an approximate global symmetry. It is an approximate symmetry since we are doing a $\frac{1}{\al^\prime E^2}$ expansion for all (tree-level) scattering amplitudes, and it is a global symmetry since we compare scattering amplitudes among independent degrees of freedom. Thus, one might be curious about the connection between the HESS and the infinite target-space gauge symmetry \cite{Witten:1985cc, Kao:2002us, Chan:2005qd}, and wonder that how it is possible to derive HESS from the decoupling of high-energy zero-norm states. For a detailed discussion, see \cite{Chan:2005ne,Chan:2005ji}. Here we try to provide a physical analogy. First of all, it is widely believed that all the massive string excitations gain their masses through a higher-spin generalization of the Higgs mechanism, in the same way as the mass generation of vector bosons in the electroweak theory. However, there seems to be some difference between these two cases. In the field theoretical context, the Higgs mechanism is facilitated by the introduction of tachyonic scalar particles to the massless gauge theory. In low-energy physics, it is more appropriate to identify the would-be Goldstone bosons (the field quanta after shifting the tachyonic scalar fields to the true vacuum) as the longitudinal degrees of freedom for the massive vector bosons. Nevertheless, in the high-energy limit ($E\gg M_w$), the advantage of such an identification is diminished. Instead, one can simply study the interactions among massless scalars with massless gauge bosons with finite Yukawa couplings and then treat $M_w/E$ corrections as perturbations in the scattering processes \cite{P}. In this view, the celebrated equivalence theorem for the gauge theory with spontaneous broken symmetry \cite{Cornwall:1974km} simply translates into the following: "The high-energy symmetry among gauge bosons scatterings in a theory endowed with spontaneous broken symmetry is nothing but a reflection of the global symmetry of the tachyonic scalar particles.".
\par
It is important to realize that, the global symmetry of the tachyonic scalar particles in field theoretical models is in principle independent from the gauge symmetry. While the tachyonic scalar particles must carry non-abelian charges to couple to gauge fields, one can impose additional global symmetry and specify suitable representation for the tachyonic scalar particles under this independent global symmetry. If we apply the same idea to the case of string theory, viewed as a higher-spin gauge theory with a super-Higgs mechanism, there are two immediate questions:
\begin{enumerate}
  \item[(1)] How to identify the would-be Glodstone bosons (presumably these will consist of a whole tower of particles with arbitrary spin)?
  \item[(2)] Modulo the issue of open-closed string duality, it seems that we know better about the gauge symmetry (e.g. in string field theory formulation), but not the global symmetry of string theory. Put it differently, what is the spontaneous symmetry breaking mechanism in string theory? Is the string gauge symmetry so powerful and restrictive such that there is only one way to incorporate spontaneous symmetry breaking?
\end{enumerate}
\par
It is clear that that the reason that we do not have good answers to the questions above is that we do not have a formulation of string theory in the most symmetry vacuum \cite{Witten:1988sy}. However, one can still try to circumvent this difficult by taking a different strategy. Following the old wisdom, it is natural to probe the string dynamics in the high-energy limits, such that through a similar mechanism as equivalence theorem, one can probe the global symmetry with the high-energy scattering amplitudes among massive gauge bosons.
\par
Once we pursue this idea further, there are still some conceptual problems that could cause confusion. Since in most practical calculations of stringy dynamics, we first choose a particular conformal invariant background and study the particle spectrum, the scattering amplitudes are calculated perturbatively (e.g. in string coupling constant $g$). In general, the particle spectrum depends on the space-time background, hence we expect the symmetry pattern among high-energy stringy scattering amplitudes should vary as we consider different space-time background \cite{Chan:2006qf,Chan:2006pf,Lee:2006fp,Lee:2007cx,Lee:2007dp,Lee:2008ba}. Furthermore, there are different kinematic limits (fixed-angle v.s. Regge) one can take in studying the high-energy stringy symmetry \cite{Ko:2008re,Ko:2008ft}. In view of these, one may raise a natural question: in what sense do these different symmetry patterns really teach us anything about the nature of string symmetry? In other words, is there a universal symmetry principle underlies these background (kinematic) dependent data?
\par
This paper is an attempt to provide some hints to the question above. Using a simple solvable string theory model, e.g. bosonic open string theory in a light-like linear dilaton background, we illustrate the concept of a universal symmetry in string theory. That is, suppose that there exists a Lie algebra structure for the global symmetry (presumably a symmetry of infinite dimension), all the conformal invariant background data (moduli parameters) are encoded in the structure constants of this infinite symmetry. Any path connecting two fixed-points in the moduli space will induce a spectral flow of particle spectrum in string theory, and we should expect a deformation of stringy symmetry from one fixed point to another. Such a symmetry deformation is nothing but a string theory generalization of group contraction, as analogous to that flattening a sphere into a plane leads to deformation of the isometry group from SO(3) to E(2).
\par

This paper is organized as follows: A brief review and summary of previous results related to this paper is given in section II. 
We also establish our notations and conventions in this section. Section III focuses on the study of covariant spectrum of the bosonic open string theory in the light-like linear dilaton background. Here we emphasize the idea of spectral flow for the physical state solutions to the Virasoro constraints. Our main results of HESS of bosonic open string theory in the light-like linear dilaton background are presented in section IV. A new fixed-point in the moduli space is identified and two sets of replacement rules for HESS at two fixed-points in the moduli space are compared. We summarize the main points of our findings in section V, emphasizing the idea of a universal stringy symmetry and its connection with a background independent formulation of string theory. Finally, we conclude by listing some future works to be done and possible speculations. 

%% file: II-A.tex
Since our starting point is very similar to that of \cite{Ho:2007ar}, we shall follow the notations in \cite{Ho:2007ar} closely. The Polyakov action for the bosonic open string theory in the linear dilaton background is given by
\beqa
\notag S&=&\frac{1}{4\pi\al^\prime}\int_\Si d^2 \si \sqrt{g}g^{ab} \partial_aX(\si)\cdot\partial_bX(\si)
+\frac{1}{4\pi}\int_\Si d^2\si \sqrt{g}R(\si)V\cdot X(\si)\\
&&+\frac{1}{2\pi}\int_{\partial\Si}ds\ka(\xi)V\cdot X(\xi),
\eeqa
here $R(\si)$ is Ricci scalar of the world-sheet $\Si$ and $\ka(\xi)$ is the geodesic curvature along the boundary of the world-sheet $\partial \Si$.
From this we can extract the energy-momentum tensor,
\beqa
T_{zz}=-\frac{1}{\al^\prime}:\partial X\cdot\partial X:+V\cdot\partial^2X.
\eeqa
The string coordinates in the oscillator representation are
\beqa
X^\mu(z,\bar{z})&=&x^\mu-i\al^\prime p^\mu \ln|z|^2+i\sqrt{\frac{\al^\prime}{2}}\sum_{m=-\infty, m\neq0}^{m=\infty}\frac{\al_m^\mu}{m}(z^{-m}+\bar{z}^{-m}).
\eeqa
The Virasoro generators of the conformal transformation are defined as the Fourier modes of the energy-momentum tensor $T_{zz}$,
\beqa
\cL_m\equiv\oint dz z^{m+1}T_{zz}=\frac{1}{2}\sum_{n=-\infty}^{\infty}:\al_{m-n}\al_n:+i\sqrt{\frac{\al^\prime}{2}}(m+1)V\cdot\al_m,
\eeqa
and they satisfy the following algebra relation,
\beqa
\big[\cL_m,\cL_n\big]=\big(m-n\big)\cL_{m+n}+\frac{D+6\al^\prime V^2}{12}m\big(m^2-1\big)\de_{m+n}.
\eeqa
Notice that the central charge includes a term which is in proportion to $V^2\equiv V^\mu V_\mu$ and we can have different space-time dimension $D$ depending on the sign of $V^2$ (space-like $V^2>0$ $\Rightarrow$ $D<26$, time-like $V^2<0$ $\Rightarrow D>26$). To simplify the calculations and to avoid the complication due to the Liouville potential in the non-critical dimension, we take the dilaton gradient, $V^\mu\equiv\partial^\mu \Phi$ to be light-like and the space-time dimension is $D=26$. 

%% file: II-B.tex
The physical state spectrum of the bosonic open string theory in the linear dilaton background is defined similarly to that of flat space-time. In the oscillator representation, we solve all possible linear combinations of creation operators acting on a Fock vacuum, subject to the Virasoro constraints:
\beqa
\cL_0\ket{\Phi(k)}=\ket{\Phi(k)},\quad \mbox{and} \quad \cL_n \ket{\Phi(k)}=0,\quad n\geqslant1.
\eeqa
These constraints in general lead to the generalized on-shell condition for the center of mass momenta, and restrict the polarization tensors to be transverse and traceless. In our study, we shall focus on the physical states up to the first massive level, and we shall use a single capital letter to represent the particles. For instance, the tachyon state (T) is given as
\beqa
\ket{T(k)}\equiv\ket{0,k}, \quad\mbox{with}\quad \al^\prime k\cdot \big(k+iV\big)=1.
\eeqa
At massless level, we have a photon state (P) with polarization vector $\zeta(k)$:
\beqa
\ket{P(\zeta, k)}\equiv\zeta_\mu\al_{-1}^\mu\ket{0,k},\quad
\notag\mbox{with}&\quad& \al^\prime k\cdot\big(k+iV\big)=0, \quad \mbox{and} \\
\label{TransvC}\zeta\cdot\big(k+iV\big)&=&0.
\eeqa
Finally, at the first massive level (M), we have a tensor particle with spin-two, and it is written as
\beqa
\ket{M(\epsilon_{\mu\nu},k)}=\big(\epsilon_{\mu\nu}\al^\mu_{-1}\al^\nu_{-1}+\epsilon_{\mu}\al^\mu_{-2}\big)\ket{0,k},\quad
\mbox{with}\quad \al^\prime k\cdot\big(k+iV\big)=-1.
\eeqa
The $\cL_{1}$ conditions, $\cL_{1}\ket{M(\epsilon_{\mu\nu},k)}=0$, gives
\beqa
\label{L1C}\sqrt{2\al^\prime}\epsilon_{\mu\nu}\big(k^\nu+iV^\nu\big)+\epsilon_\mu=0,
\eeqa
and the $\cL_2$ conditions, $\cL_{2}\ket{M(\epsilon_{\mu\nu},k)}=0$, gives
\beqa
\label{L2C}\epsilon_{\mu\nu}\eta^{\mu\nu}+\sqrt{2\al^\prime}\epsilon_\mu\big(2k^\mu+3iV^\mu\big)=0.
\eeqa
Substituting $\epsilon_\mu$ from Eq.\eqref{L1C} to Eq.\eqref{L2C}, we get
\beqa
\label{Vra}2\al^\prime\epsilon_{\mu\nu}\big(k^\mu+iV^\mu\big)\big(2k^\nu+3iV^\nu\big)-\epsilon_{\mu\nu}\eta^{\mu\nu}=0.
\eeqa
Note that all of these relations contain explicit dependence on the linear dilaton gradient $V^\mu$, and one can verify that as $V^\mu$ goes to zero, we recover all the previous results on physical spectrum for bosonic open string in flat space-time \cite{Chan:2005qd}. It is then natural to treat $V^\mu$ as a moduli parameter and identify the solutions to the Virasoro constraints as a spectral flow. In addition to the interpretation of physical spectrum deformation (as a function of $V^\mu$), it is also crucial to emphasize that the inner product in the one string Fock space in the linear dilaton background is also deformed. Here we follow the prescription in \cite{Ho:2007ar} and define the inner product for the center of mass degree of freedom of any stringy excitation,
\beqas
\braket{k^\prime}{k}\equiv\big(2\pi\big)^D\de^{(D)}\big(k^{\prime\ast}-k-iV\big).
\eeqas
One should be cautious about the definition of zero-norm states with respect to the deformed inner product and check that the gauge invariance (decoupling of the zero-norm states) is maintained in the presence of a linear dilaton background \cite{Chan:2009xx}. 

%% file: II-C.tex
Functional integration method has been applied to the calculations of scattering amplitudes of tachyon and photon states in the presence of a light-like linear dilaton background \cite{Ho:2007ar}.
We approached and extended the similar calculations based on the operator methods \cite{Green:1987sp,Chan:2009xx,Chan:2009vx}. The explicit forms of vertex operators of physical states are shown to satisfy the conformal algebra \cite{Green:1987sp},
\beqa
\Big[\cL_m, \cV(\tau)\Big]=e^{im\tau}\Big(-i\frac{d}{d\tau}+m\Big)\cV(\tau).
\eeqa
Here we list the results for normal-ordered vertex operators of tachyon(T), photon(P) and the massive tensor(M),
\beqa
  \label{vetx1}\ket{T(k)}&\Rightarrow& e^{-\al^\prime k\cdot V\tau}:e^{ik\cdot X}: \\
  \label{vetx2} \ket{P(\zeta,k)}&\Rightarrow& \frac{\zeta\cdot \big(\dot X+i\al^\prime V\big)}{\sqrt{2\al^\prime}}e^{-\al^\prime k\cdot V\tau}:e^{ik\cdot X}:  \\
  \label{vetx3}\ket{M(\epsilon_{\mu\nu},k)}&\Rightarrow&\Big[\frac{\epsilon_{\mu\nu}}{2\al^\prime}\big(\dot X^\mu+i\al^\prime V^\mu\big)\big(\dot X^\nu+i\al^\prime V^\nu\big)-\frac{i\epsilon}{\sqrt{2\al^\prime}}\cdot \ddot X\Big]e^{-\al^\prime k\cdot V\tau}:e^{ik\cdot X}: .
\eeqa
Given these explicit forms, it is straightforward to obtain any stringy scattering amplitudes. Interested readers are invited to consult our paper in \cite{Chan:2009xx}, and we shall use the results from \cite{Chan:2009xx} directly in section VI. 

%% file: III-A.tex
Using the generators of the Virasoro algebra for the open string theory, one can construct zero-norm states (ZNS), which generate stringy gauge symmetry \cite{Green:1987sp, Lee:1989rc, Lee:1994wp}. At the first massive level ($\al^\prime m^2=1$), we have two types of zero-norm states:

\beqa
\mbox{type I vector ZNS},\qquad \cL_{-1}\ket{\chi}\equiv\cL_{-1}\big(\epsilon_\mu\al_{-1}^\mu\big)\ket{0,k},
\eeqa
where the "seed state" $\ket\chi\equiv\epsilon\cdot\al_{-1}\ket{0,k}$, satisfies the following conditions,
\beqa
  \cL_{0}\ket{\chi}=0 & \Rightarrow & \al^\prime k\cdot\big(k+iV\big)=-1 \quad\mbox{(on-shell condition)},\\
  \label{L1}\cL_{1}\ket{\chi}=0 & \Rightarrow & \epsilon\cdot \big(k+iV\big)=0 \qquad\mbox{(transverse condition)},
\eeqa
and $\cL_{2}\ket{\chi}=0$ holds automatically.
If we use the oscillator representation of Virasoro generator,
\beqas
\cL_{-1}\sim\al_{-1}\al_0+\al_{-2}\al_1,
\eeqas
we can read out the polarizations of the vector zero-norm states,
\beqas
\cL_{-1}\ket \chi=\big(\epsilon_{\mu\nu}\al^\mu_{-1}\al^\nu_{-1}+\epsilon_\mu\al^\mu_{-2}\big)\ket{0,k},
\eeqas
where $\displaystyle\epsilon_{\mu\nu}=\sqrt\frac{\al^\prime}{2}\big(\epsilon_\mu k_\nu+\epsilon_\nu k_\mu\big)$.
\par
In order to solve for the polarization vector $\epsilon_\mu$, Eq.\eqref{L1}, it is useful to express all momenta in the helicity basis.
In the presence of a linear dilaton background, the momentum of the first massive ($\al^\prime m_2^2=1$) spin-two particle is chosen as
\beqas
k\equiv \big(E-i\frac{V_0}{2}, \mathrm{k}+i\frac{V_0}{2},0\big),
\eeqas
and the helicity basis for the planar scatterings consists of the following vectors \cite{Chan:2005qd}:
\beqa
\notag e^P&\equiv&\sqrt{\al^\prime}\big(E, \mathrm{k},0\big),\\
\label{a41}e^L&\equiv&\sqrt{\al^\prime}\big(\mathrm{k}, E,0\big),\\
\notag e^T&\equiv&\sqrt{\al^\prime}\big(0,0,1\big).
\eeqa
It is clear that these orthonormal vectors satisfy the completeness relation:
\beqas
e^\al\cdot e^\be=\eta^{\al\be},\qquad \sum_{\al}\big(e^\al\big)_\mu\big(e_\al\big)_\nu=\eta_{\mu\nu},\qquad \al, \be=P,L,T_i.
\eeqas
For later convenience, it is useful to decompose the following vectors in terms of the helicity basis,
\beqa
\notag k&=&\frac{1}{\sqrt{\al^\prime}}\big(a_1e^P+b_1e^L\big),\\
\label{Kimdecp}k+iV&=&\frac{1}{\sqrt{\al^\prime}}\big(a_2e^P+b_2e^L\big),\\
\notag k+\frac{3i}{2}V&=&\frac{1}{\sqrt{\al^\prime}}\big(a_3e^P+b_3e^L\big).
\eeqa
The expansion coefficients are:
\beqas
v\equiv \al^\prime(E+\mathrm{k})V_0,
\eeqas
\beqas
\begin{array}{lc}
  \displaystyle a_1=1-\frac{i}{2}v, & \displaystyle b_1=\frac{i}{2}v; \\
  \displaystyle a_2=1+\frac{i}{2}v, & \displaystyle b_2=-\frac{i}{2}v; \\
  \displaystyle a_3=1+iv, & \displaystyle b_3=-iv.
\end{array}
\eeqas
From Eq.\eqref{L1}, it is clear that we have 25 solutions for the transverse polarization vector, $\epsilon_\mu$. Specifically, we shall use the following vectors
\beqas
\epsilon(L)\propto b_2e^P+a_2e^L,\quad\mbox{and}\quad\epsilon(T_i)\propto e^{T_i}.
\eeqas
to construct all type I vector zero-norm states:
\begin{enumerate}
  \item[Case 1:]
  \beqa
  \notag\epsilon_\mu{(L)}&\equiv& \sqrt{2}\big(b_2e^P_\mu+a_2e^L_\mu\big),\\
  \notag\Rightarrow \epsilon_{\mu\nu}(L)&=& 2a_1b_2e^P_\mu e^P_\nu+2a_2b_1e^L_\mu e^L_\nu+\big(a_1a_2+b_1b_2\big)\big(e^P_\mu e^L_\nu+e^P_\nu e^L_\mu\big).\\
  \notag\ket{ZNS_I(L)}&=&\Big[\epsilon_{\mu\nu}(L)\al^\mu_{-1}\al_{-1}^\nu+\epsilon_\mu(L)\al^\mu_{-2}\Big]\ket{0,k}\\
  \notag&=&\left[\begin{split}&2a_1b_2\al_{-1}^P\al_{-1}^P+2a_2b_1\al_{-1}^L\al_{-1}^L\\
  &+2\big(a_1a_2+b_1b_2\big)\al_{-1}^P \notag\al_{-1}^L+\sqrt{2}\big(b_2\al_{-2}^P+a_2\al_{-2}^L\big)\end{split}\right]\ket{k,0}\\
  \label{ZNSI1}&=&
\left[\begin{split}&\big(-iv-\frac{v^2}{2}\big)\al_{-1}^P\al_{-1}^P+\big(iv-\frac{v^2}{2}\big)\al_{-1}^L\al_{-1}^L\\&+\big(2+ v^2\big)\al_{-1}^P \al_{-1}^L-\frac{iv}{\sqrt 2} \al^P_{-2}+\Big(\sqrt 2+\frac{iv}{\sqrt 2} \Big)\al^L_{-2}\end{split}\right]\ket{k,0}.
  \eeqa
  \item[Case 2:]
  \beqa
  \notag\epsilon_\mu(T_i)&\equiv& \frac{\sqrt 2}{a_1}e^{T_i}_\mu,\\
  \notag\Rightarrow \qquad\epsilon_{\mu\nu}(T_i)&=&\big(e^{T_i}_\mu e^P_\nu+e^{T_i}_\nu e^P_\mu\big)+\frac{b_1}{a_1} \big(e^{T_i}_\mu e^L_\nu+e^{T_i}_\nu e^L_\mu\big).\\
  \notag\ket{ZNS_{I}(T_i)}&=&\Big[\epsilon_{\mu\nu}(T_i)\al^\mu_{-1}\al_{-1}^\nu+\epsilon_\mu(T_i)\al^\mu_{-2}\Big]\ket{0,k}\\
  \notag&=&\sum_{i=1}^{24}u_{PT_i}\Big[2\al_{-1}^P\al_{-1}^{T_i}+2\frac{b_1}{a_1}\al_{-1}^L\al_{-1}^{T_i}+\frac{\sqrt2}{a_1}\al_{-2}^{T_i}\Big]\ket{0,k}\\
  \label{ZNSI2}&=&\sum_{i=1}^{24}u_{PT_i}\bigg[2\al_{-1}^P\al_{-1}^{T_i}+\frac{iv}{2-iv}\al_{-1}^L\al_{-1}^{T_i}+\frac{2\sqrt 2}{2-iv}\al_{-2}^{T_i}\bigg]\ket{0,k}
  .\eeqa
\end{enumerate}
\par The type II ZNS at the first massive level can be calculated by the same formula as that in the flat space-time. We have
\beqa
\notag\ket{\varphi(k)}&\equiv&\big(2\cL_{-2}+3\cL_{-1}^2\big)\ket{0,k}\\
\label{ZNSII}&=&\Big[\epsilon_{\mu\nu}\al_{-1}^\mu\al_{-1}^\nu+\epsilon_\mu\al_{-2}^\mu\Big]\ket{0,k},\\
\notag\mbox{where}\qquad \epsilon_{\mu\nu}&=&6\al^\prime k_\mu k_\nu+\eta_{\mu\nu},\\
\notag\mbox{and}\qquad\quad \epsilon_\mu&=&\sqrt{2\al^\prime}\big(5k_\mu-iV_\mu\big).
\eeqa
One can check that the normalization of $\ket \varphi$ is
\beqas
\braket{\varphi(k^\prime)}{\varphi(k)}&=&2\Big[\epsilon_{\mu\nu}(k)\epsilon^{\ast\mu\nu}(k^\prime)
+\epsilon_\mu(k)\epsilon^{\ast\mu}(k^\prime)\Big]\de(k^{\prime\ast}-k-iV)\\
&=&4\left[\begin{split}&18\al^{\prime2}\big(k\cdot k^{\prime\ast}\big)^2+3\al^\prime k^2+3\al^\prime\big(k^{\prime\ast}\big)^2\\
&+25\al^\prime\big(k\cdot k^{\prime\ast}\big)+5i\al^\prime\big(k-k^{\prime\ast}\big)V+\frac{D^2}{2}+\al^\prime V^2\end{split}\right]\de(k^{\prime\ast}-k-iV).
\eeqas
Substitute the on-shell condition $\al^\prime k\cdot\big(k+iV\big)=-1$ and $\al^\prime k^\ast\cdot\big(k^\ast-iV\big)=-1$, one can verify that $\ket{\varphi(k)}$ is indeed a zero-norm state for all $D$ and $V^\mu$.
Expanding all the Virasoro generators in Eq.\eqref{ZNSII}, we get
\beqas
\epsilon_{\mu\nu}&=&6\al^\prime k_\mu k_\nu+\eta_{\mu\nu}\\
    &=&\big(6a_1^2-1\big)e^P_\mu e^P_\nu+\big(6b_1^2+1\big)e^L_\mu e^L_\nu+
  \sum_{i=1}^{24}e^{T_i}_\mu e^{T_i}_\nu+6a_1 b_1\big(e^P_\mu e^L_\nu+e^P_\nu e^L_\mu\big),
\eeqas
and
\beqas
\epsilon_\mu&=&-\sqrt{2\al^\prime}\epsilon_{\mu\nu}\big(k^\nu+iV^\nu\big)\\
&=&\sqrt{2}\Big[6\al^\prime a_1\big(a_1a_2-b_1b_2\big)-a_2\Big]e^P_\mu+\sqrt{2}\Big[6\al^\prime b_1\big(a_1a_2-b_1b_2\big)-b_2\Big]e^L_\mu.
\eeqas
Putting all these ingredients together, we get
\beqa
\notag\ket{ZNS_{II}}&=&\Big(\epsilon_{\mu\nu}\al^\mu_{-1}\al_{-1}^\nu+\epsilon_\mu\al^\mu_{-2}\Big)\ket{0,k}\\
\notag&=&\left\{\begin{split}\Big[\big(6a_1^2-1\big)\al^P_{-1}\al^P_{-1}+\big(6 b_1^2+1\big)\al^L_{-1}\al^L_{-1}+\sum_{i=1}^{24}\al_{-1}^{T_i}\al_{-1}^{T_i}
+12a_1b_1\al_{-1}^{P}\al_{-1}^{L}\Big]\\+\sqrt{2}\Big[6a_1\big(a_1a_2-b_1b_2\big)-a_2\Big]\al^P_{-2}+
\sqrt{2}\Big[6 b_1\big(a_1a_2-b_1b_2\big)-b_2\Big]\al^L_{-2}\end{split}\right\}\ket{0,k}\\
\label{ZNSII2}&=&\left[\begin{split}&\big(5-6iv-\frac{3v^2}{2} \big)\al_{-1}^P \al_{-1}^P+\big(1-\frac{3v^2}{2} \big)\al_{-1}^L \al_{-1}^L+\sum_{i=1}^{24}\al_{-1}^{T_i} \al_{-1}^{T_i}\\
&+\big(6iv+3 v^2\big)\al_{-1}^P \al_{-1}^L
+\Big(\frac{10}{\sqrt 2}-\frac{7iv}{\sqrt 2}\Big)\al^P_{-2}+\frac{7iv}{\sqrt 2}\al_{-2}^L
\end{split}\right]\ket{0,k}.
\eeqa 

%% file: III-B.tex
After transforming the spin-two polarization tensor $\epsilon_{\mu\nu}$ into helicity basis,
\beqa
\epsilon_{\mu\nu}\equiv\sum_{\al,\be}u_{\al\be}\big(e^\al\big)_\mu\big(e^\be\big)_\nu,\quad \al,\be=P,L,T_i,
\eeqa
the Virasoro constraint, Eq.\eqref{Vra}, becomes
\beqa
\notag&&4\al^\prime\epsilon_{\mu\nu}\big(k_2^\mu+iV^\mu\big)\big(k_2^\nu+\frac{3}{2}iV^\nu\big)=\epsilon_{\mu\nu}\eta^{\mu\nu}\\
\notag&\Rightarrow&4\epsilon_{\mu\nu}\Big[a_2(e^P)^\mu+b_2(e^L)^\mu\Big]\Big[a_3(e^P)^\nu+b_3(e^L)^\nu\Big]=-u^{PP}+u^{LL}+\sum_{i=1}^{24}u^{T_iT_i}\\
&\Rightarrow&
\big(4 a_2a_3+1\big)u^{PP}+(4 b_2b_3-1\big)u^{LL}+\big(4 a_2b_3+4 a_3b_2\big)u^{PL}-\sum_{i=1}^{24}u^{T_iT_i}=0.
\eeqa
On the other hand, $\cL_1$ condition implies that the polarization vector can be derived as a projection of the spin-two tensor $\epsilon_{\mu\nu}$,
\beqa
\epsilon_\mu&=&-\sqrt{2\al^\prime}\epsilon_{\mu\nu}\big(k^\nu+iV^\nu\big)\\
&=&-\sqrt{2}\Big[\sum_{A,B=P,L,T_i}u_{AB}e^A_\mu e^B_\nu\Big]\Big[a_2\big(e^P\big)^\mu+b_2\big(e^L\big)^\mu\Big]\\
\notag&=&-\sqrt{2}\bigg(-a_2u_{PP}+b_2u_{PL}\bigg)e^P_\mu-\sqrt{2}\bigg(-a_2u_{PL}+b_2u_{LL}\bigg)e^L_\mu
\\&&-\sqrt{2}\sum_{i=1}^{24}\bigg(-a_2u_{PT_i}+b_2u_{LT_i}\bigg)e^{T_i}_\mu\\
\notag&=&\bigg[\Big(\sqrt 2+\frac{iv}{\sqrt2}\Big)u_{PP}+\Big(\frac{iv}{\sqrt2}\Big)u_{PL}\bigg]e^P_\mu+\bigg[\Big(\sqrt 2+\frac{iv}{\sqrt2}\Big)u_{PL}+\Big(\frac{iv}{\sqrt2}\Big)u_{LL}\bigg]e^L_\mu
\\&&+\sum_{i=1}^{24}\bigg[\Big(\sqrt 2+\frac{iv}{\sqrt2}\Big)u_{PT_i}+\Big(\frac{iv}{\sqrt2}\Big)u_{LT_i}\bigg]e^{T_i}_\mu.
\eeqa
The solutions of positive-norm states at $\al^\prime m^2=1$ level in a linear dilaton background are given by
\beqa
\notag\ket{PNS}_1&=&\sum_{i=1}^{24}\Big[\al^{L}_{-1}\al^{L}_{-1}+\big(4 b_2b_3-1\big)\al^{T_i}_{-1}\al^{T_i}_{-1}-\sqrt 2b_2\al^L_{-2}\Big]\ket{0,k}\\
&=&\sum_{i=1}^{24}\Big[\al^{L}_{-1}\al^{L}_{-1}-\big(1+2 v^2\big)\al^{T_i}_{-1}\al^{T_i}_{-1}+\frac{iv}{\sqrt2}\al^{L}_{-2}\Big]\ket{0,k},\\
\notag\ket{PNS}_2&=&\sum_{i=1}^{24}w_i\Big(\al^{L}_{-1}\al^{T_i}_{-1}-\frac{b_2}{\sqrt 2}\al^{T_i}_{-2}\Big)\ket{0,k}\\
&=&\sum_{i=1}^{24}w_i\Big(\al^{L}_{-1}\al^{T_i}_{-1}+\frac{iv}{2\sqrt2}\al_{-2}^{T_i}\Big)\ket{0,k},\\
\ket{PNS}_3&=&\sum_{i,j}\Big(u_{T_iT_j}-\frac{\de_{ij}}{24}\sum_{\ell=1}^{24}u_{T_\ell T_\ell}\Big)\al^{T_i}_{-1}\al^{T_j}_{-1}\ket{0,k}.
\eeqa
\par In the later calculations of stringy scattering amplitudes, we should use normalized positive-norm states (or, equivalently, vertex operators) as inputs. This is important in comparing the high-energy limits of different physical string scattering amplitudes as a manifestation of high-energy stringy symmetry. For this reason, we choose to represent the positive-norm states for the spin-two particles as
\beqa
\notag\ket{PNS}_1\quad\rightarrow\quad\ket{M(LL)}&\equiv&\frac{1}{\sqrt{8v^4+9v^2+4}}\Big[\al^{L}_{-1}\al^{L}_{-1}-\big(1+2 v^2\big)\al^{T}_{-1}\al^{T}_{-1}+\frac{iv}{\sqrt2}\al^{L}_{-2}\Big]\ket{0,k},\\
\label{PNSnomo}\ket{PNS}_2\quad\rightarrow\quad\ket{M(LT)}&\equiv&\frac{2}{\sqrt{v^2+4}}\Big(\al^{L}_{-1}\al^{T}_{-1}+\frac{iv}{2\sqrt2}\al_{-2}^{T}\Big)\ket{0,k},\\
\notag\ket{PNS}_3\quad\rightarrow\quad\ket{M(TT)}&\equiv&\frac{1}{\sqrt 2}\al^{T}_{-1}\al^{T}_{-1}\ket{0,k}.
\eeqa
It is interesting to notice that at this massive level, if we tune the moduli parameter $v$ from zero to infinity, the spectrum apparently degenerated,
\beqas
\ket{M(LL), v\rightarrow\infty}\propto\al^{T}_{-1}\al^{T}_{-1}\ket{0,k}\propto\ket{M(TT),v\rightarrow\infty}.
\eeqas
However, we shall see later that, actually the contributions from $\al^{L}_{-1}\al^{L}_{-1}$ piece of $\ket{M(LL)}$ state to the high-energy stringy scattering amplitudes relative to those of $\al^{T}_{-1}\al^{T}_{-1}$ piece of $\ket{M(LL)}$ is of order $v^2$. Hence, it has non-vanishing contribution to the scattering amplitudes and one should be careful in interpreting the degenerate spectrum at $v\rightarrow\infty$ limit. 

%% file: III-C.tex
Having identified all the independent basis states of the covariant spectrum at $\al^\prime m^2=1$ in a light-like linear dilaton background, we can now write down the most general decomposition and find the solution to the Virasoro constraints at $\al^\prime m^2=1$. Introducing a new sets of expansion coefficients \{$x$, $y$, $z$, $w_i$\}, we have
\beqa
\notag&&\Big[\epsilon_{\mu\nu}\al^\mu_{-1}\al^\nu_{-1}+\epsilon_{\mu}\al^{\mu}_{-2}\Big]\ket{0,k}\\
\label{A1} &=&x\ket{ZNS_{II}}+y\ket{ZNS_I(L)}+\ket{ZNS_I(T)}+z\ket{PNS}_1+\ket{PNS}_2+\ket{PNS}_3.
\eeqa

Comparing both sides of Eq.\eqref{A1}, we get
\beqas
u_{PP}&=&\big(6a_1^2-1\big)x+2a_1b_2y,\\
u_{PL}&=&6a_1b_1x+\big(a_1a_2+b_1b_2\big)y,\\
u_{LL}&=&\big(6b_1^2+1\big)x+2 a_2b_1y+24z,\\
u_{LT_i}&=&\frac{b_1}{a_1}u_{PT_i}+\frac{1}{2}w_i,\\
\frac{1}{24}\sum_{\ell=1}^{24}u_{T_\ell T_\ell}&=&x+\big(4 b_2b_3-1\big)z.
\eeqas
Solving $x$, $y$, $z$, $w_i$ in terms of $u_{\al\be}$, we get
\beqa
\notag x&=&\de\Big[\big(a_1a_2+b_1b_2\big)u_{PP}-2 \big(a_1b_2\big)u_{PL}\Big],\\
\notag y&=&\de\Big[-\big(6a_1b_1\big)u_{PP}+\big(6a_1^2-1\big)u_{PL}\Big],\\
\label{A2}24z&=&-\de\Big[a_1a_2\big(1-6 b_1^2\big)+b_1b_2\big(1+6 b_1^2\big)\Big]u_{PP},\\
\notag&&+2\de\Big[a_1b_2\big(1+6 b_1^2\big)+a_2b_1\big(1-6 a_1^2\big)\Big]u_{PL}+u_{LL},\\
\notag w_i&=&-2\frac{b_1}{a_1}u_{PT_i}+2u_{LT_i},\\
\notag\de^{-1}&\equiv&6 a_1^2\big(a_1a_2-b_1b_2\big)-\big(a_1a_2+b_1b_2\big).
\eeqa
Substituting the solutions of ($a_i$, $b_i$) in Eqs.\eqref{A2}, we have
\beqas
x&=&\de\Big[\big(1+\frac{v^2}{2}\big)u_{PP}+\big(iv+\frac{v^2}{2}\big)u_{PL}\Big],\\
y&=&\de\Big[\big(-3iv-\frac{3v^2}{2} \big)u_{PP}+\big(5-6iv-\frac{3v^2}{2}\big)u_{PL}\Big],\\
24z&=&-\de\big(1+2 v^2\big) u_{PP}-\de\big(6iv+4v^2\big)u_{PL}+u_{LL},\\
w_i&=&-\frac{2iv}{2-iv}u_{PT_i}+2u_{LT_i},\\
\de^{-1}&=&iv-\frac{3v^2}{4}.
\eeqas

%% file: IV-A-1.tex
Our choice of kinematic variables for three-point functions in the light-like linear dilaton background is based on the following diagram (subscripts for momenta or polarizations denote labels of particles):
\begin{figure}[h]
\begin{center}\includegraphics[width=0.8\textwidth]{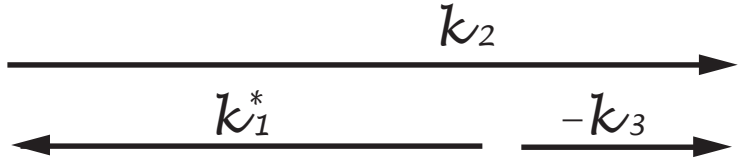}\end{center}
\caption{Kinematic configuration for three-point functions}
\end{figure}
~\\
One can imagine a heavy particle moving down the slope of a linear dilaton background with momentum $k_2$ and decaying into two particles. While $-k_3$ stands for the momentum of the right-moving remanet, we have another particle carrying momentum $k_1$ moving toward left. In terms of components, we have:
\begin{enumerate}
  \item[(i)] light-like dilaton gradient,
  $V^\mu\equiv\partial^\mu\Phi$, $V\equiv\big(V_0,V_1,V_2\big)$.
  \item[(ii)] momenta of the second and the third particles,
  \beqas
  \begin{array}{ll}
    \displaystyle k_2\equiv\big(E_2-\frac{iV_0}{2},\mathrm{k}_2-\frac{iV_1}{2}, -\frac{iV_2}{2}\big),& \quad \displaystyle k_2\cdot\big(k_2+iV\big)=-m_2^2\quad\Rightarrow\quad E_2^2-\mathrm{k}_2^2=m_2^2. \\
    &\\
    \displaystyle k_3\equiv\big(-E_3-\frac{iV_0}{2},-\mathrm{k}_3-\frac{iV_1}{2}, -\frac{iV_2}{2}\big), & \quad k_3\cdot\big(k_3+iV\big)=-m_3^2\quad\Rightarrow\quad E_3^2-\mathrm{k}_3^2=m_3^2.
  \end{array}
  \eeqas
  \item[(iii)] momentum of the first particle,
  \beqa
  \label{3EMC}\hspace{-5mm}k_1^\ast=k_2+k_3+iV=\big(E_2-E_3, \mathrm{k}_2-\mathrm{k}_3,0\big)=k_1.
  \eeqa
  Here we have imposed the momentum conservation to obtain $k_1$.
  \item[(iv)] on-shell condition for $k_1$,
  \beqa
  \label{OS1}k_1\cdot\big(k_1+iV\big)=-m_1^2.
  \eeqa
\end{enumerate}
If we substitute the explicit form, Eq.\eqref{3EMC}, into the on-shell condition for $k_1$, Eq.\eqref{OS1}, we obtain
\beqa
\label{DetofV1}2\big(E_2E_3-\mathrm{k}_2\mathrm{k}_3\big)=m_2^2+m_3^2-m_1^2,
\eeqa
and
\beqa
\label{DetofV2} V_0\big(E_2-E_3\big)=V_1\big(\mathrm{k}_2-\mathrm{k}_3\big),
\eeqa
from the real and imaginary parts of Eq.\eqref{OS1}, respectively.
Since we have assumed that the dilaton gradient is light-like, $V^2_2=V_0^2-V_1^2$, it turns out that, based on Eq.\eqref{DetofV2},
\beqa
\label{DetofV3}V_2^2=-\frac{m_1^2}{\big(E_2-E_3\big)^2}V_1^2,
\eeqa
and we conclude that both $V_2$ and $m_1$ must vanish and $V_0^2=V_1^2$.\par
One can solve all kinematic variables in terms of $\mathrm{k}_2$, in particular, we get (for $m_1=0$),
\beqas
E_2&=&\sqrt{m_2^2+\mathrm{k}_2^2},\\
\mathrm{k}_3&=&\frac{\big(m_2^2+m_3^2\big)\mathrm{k}_2+\big(m_2^2-m_3^2\big)E_2}{2m_2^2},\\
E_3&=&\frac{\big(m_2^2+m_3^2\big)E_2+\big(m_2^2-m_3^2\big)\mathrm{k}_2}{2m_2^2}.
\eeqas
One can check that these solutions are consistent with the on-shell conditions for $k_1$, Eqs.\eqref{DetofV1}, \eqref{DetofV2}. In addition, one can verify that $E_2-E_3=\mathrm{k}_3-\mathrm{k}_2$. Hence we set $V_1=-V_0$.
\par In later calculations of the three-point functions involving massive spin-two particles in the helicity representation Eq.\eqref{a41}, it is useful to make the following momentum decomposition:
\beqa
\label{dec1}\sqrt{\al^\prime}k_1&=&\Big(-\frac{(m_2^2-m_3^2)(\mathrm{k}_2-E_2)}{2m^3_2}, \frac{(m_2^2-m_3^2)(\mathrm{k}_2-E_2)}{2m^3_2}, 0\Big)=a_4e^P+b_4e^L,\\
\label{dec2}\sqrt{\al^\prime}k_2&=&\sqrt{\al^\prime}\Big(E_2-\frac{iV_0}{2}, \mathrm{k}_2+\frac{iV_0}{2}, 0\Big)=a_1e^P+b_1e^L,\\
\notag \sqrt{\al^\prime}k_3&=&\Big(-\frac{(m_2^2+m_3^2)E_2+(m_2^2-m_3^2)\mathrm{k}_2}{2m_2^3}-\frac{iV_0}{2m_2}
, -\frac{(m_2^2+m_3^2)\mathrm{k}_2+(m_2^2-m_3^2)E_2}{2m_2^3}+\frac{iV_0}{2m_2}, 0\Big)\\
\label{dec3}&=&a_5e^P+b_5e^L,
\eeqa
where
\beqas
\begin{array}{ll}
 \displaystyle a_4=\frac{\al^{\prime}\be_{23}}{2}, \quad \displaystyle b_4=-\frac{\al^{\prime}\be_{23}}{2}, \quad
\displaystyle a_5=-\al^{\prime}\al_{23}-\frac{iv}{2},   \quad  \displaystyle b_5=-\frac{\al^{\prime}\be_{23}}{2}+\frac{iv}{2}.
\end{array}
\eeqas
Here $\displaystyle\al_{ij}\equiv\frac{m_i^2+m_j^2}{2}$ is the average mass squared, and $\be_{ij}\equiv m_i^2-m_j^2$ is the difference between mass squared.

\par It is important to emphasize that even though a special kinematics has been chosen to study the scattering amplitudes, our goal is to provide a concrete illustration of the universality of stringy symmetry in a simple setup. To be more specific, while it is possible to have other interesting features by including more free parameters in the kinematics, here we shall take $V_0$ as the moduli parameter and extract the behavior of all stringy scattering amplitudes as $V_0$ continuously evolves from zero to infinity. 

%% file: IV-A-2.tex
To calculate the high-energy limits of various four-point functions in the light-like linear dilaton background, we choose to work with the "center of momentum" frame and define the following kinematic variables:
\begin{figure}[h]
\begin{center}\includegraphics[width=0.6\textwidth]{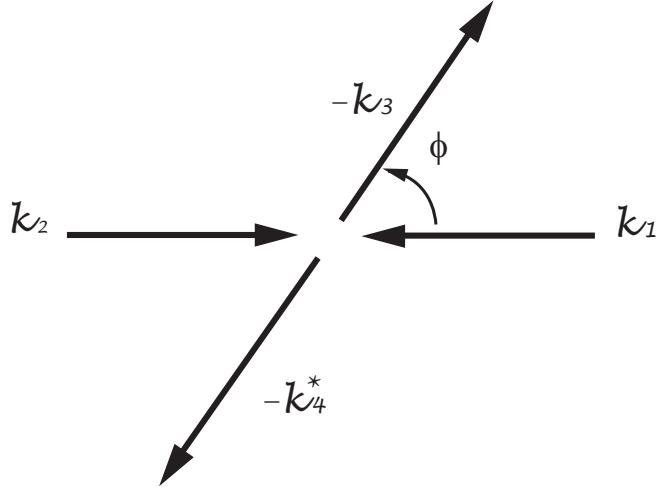}\end{center}
\caption{Kinematic configuration for four-point functions}
\end{figure}
\begin{enumerate}
  \item[(i)] light-like dilaton gradient, $V^\mu\equiv\partial^\mu\Phi$, $V\equiv\big(V_0,-V_0,0\big)$.
  \item[(ii)] momentum of the first particle (photon):
  \beqas
  &&k_1\equiv\big(E_1+\frac{iV_0}{2},-\mathrm{k}_2-\frac{iV_0}{2}, 0\big),\quad \mathrm{k}_2\geqslant 0\quad \mbox{(left-moving along x-axis)},\\
  &&k_1\cdot\big(k_1+iV\big)=-m_1^2=-E_1^2+\mathbf{k}_2^2=0.
  \eeqas
  The reason that $V_1=-V_0$ and $m_1=0$ is the same as the case of three-point functions, see Eq.\eqref{DetofV3}.
  \item[(iii)]momentum of the second particle:
  \beqas
  &&k_2\equiv\big(E_2-\frac{iV_0}{2},\mathrm{k}_2+\frac{iV_0}{2}, 0\big),\qquad \mbox{(right-moving along x-axis)}\\
  &&k_2\cdot\big(k_2+iV\big)=-m_2^2=-E_2^2+\mathbf{k}_2^2.
  \eeqas
  \item[(iv)]momentum of the third particle:
  \beqa
  && \label{Kim3}k_3\equiv\big(-E_3-\frac{iV_0}{2},-\mathrm{k}_3\cos\phi+\frac{iV_0}{2}, -\mathrm{k}_3\sin\phi\big),\quad \mbox{($\phi$ is the scattering angle)}\\
  &&\notag k_3\cdot\big(k_3+iV\big)=-m_3^2=-E_3^2+\mathbf{k}_3^2.
  \eeqa
  \item[(v)]momentum of the fourth particle:
  \beqas
  &&k_4\equiv\big(E_1+E_2-E_3-\frac{iV_0}{2},-\mathrm{k}_3\cos\phi+\frac{iV_0}{2}, -\mathrm{k}_3\sin\phi\big),\\
  &&k_4\cdot\big(k_4+iV\big)=-m_4^2=-\big(E_1+E_2-E_3\big)^2+\mathbf{k}_3^2=0.
  \eeqas
  \item[(vi)]momentum conservation: $k_4^{\ast}=k_1+k_2+k_3+iV$.
\end{enumerate}
Our choice of kinematic variables follows the conventions of previous works \cite{Chan:2004yz,Chan:2004tb,Chan:2005ne,Chan:2005zp,Chan:2005ji} on the high-energy stringy symmetry and should be considered as a minimal extension of the bosonic open string dynamics in the flat space-time. Thus, all the flat space-time results will be recovered if we set $V_0=0$, and we shall identify a new symmetry pattern as $V_0/E$ tends to infinity. 

%% file: IV-A-3.tex
In the calculations of stringy scattering amplitudes, we shall need the polarization vector for photon. Specifically,
 we take the transverse polarization vectors for photon moving along $x$-axis as
\beqas
e^T_1=e^T_2=(0, 0, 1) .
\eeqas
Given the momentum of the scattered (the third) particle $k_3$, we need to solve for the transverse condition, Eq.\eqref{TransvC}, $e^T_3\cdot (k_3+iV)=0$, for the polarization vector. Based on our kinematics setup, Eq.\eqref{Kim3}, it is easy to check that $e^T_3$ is proportional to the following vector,
\beqas
e^T_3=\frac{1}{N}\big(0,-\mathrm{k}_3\sin\phi,\mathrm{k}_3\cos\phi+\frac{iV_0}{2}\big),
\eeqas
and the normalization constant, $N$, is given as $N\equiv\sqrt{\displaystyle\mathrm{k}_3^2+\frac{V_0^2}{4}}$. 

%% file: IV-A-4.tex
Since the energy $E_i$ and momentum $\mathrm{k}_i$ variables  satisfy the same on-shell conditions as those in the flat space-time, we define the average center of momentum energy $E$, as
\beqa
E_1+E_2=2E=E_3+E_4.
\eeqa
In the high-energy limit, we have \cite{CCP}
\beqa
E_1=E+\frac{\be_{12}}{4E}
,\quad E_2=E-\frac{\be_{12}}{4E},&\quad&
E_3=E+\frac{\be_{34}}{4E},\quad E_4=E-\frac{\be_{34}}{4E};\\
\mathrm{k}_1=\mathrm{k}_2\sim E-\frac{\al_{12}}{2E},&\quad&
\mathrm{k}_3=\mathrm{k}_4\sim E-\frac{\al_{34}}{2E}.
\eeqa
\par The Mandelstam variables, in the light-like linear dilaton background, are defined as
\beqa
s&=&-\big(k_1+k_2\big)\cdot\big(k_1+k_2+iV\big)=4E^2+2iEV_0,\\
t&=&-\big(k_2+k_3\big)\cdot\big(k_2+k_3+iV\big)\\
&=&2\big(-E_2E_3+\mathrm{k}_2\mathrm{k}_3\cos\phi+\al_{23}\big)-i\big(E_2-E_3+\mathrm{k}_2-\mathrm{k}_3\cos\phi\big)V_0,\\
&\quad&s+t+u=\sum_{i=1}^4 m_i^2.
\eeqa
Some useful results  for the contractions among momenta and polarization vectors are listed in Table \ref{1}.
\begin{table}
\begin{center}
\subfigure[The first and the second particle]{\begin{tabular}{|c|c|c|}
\hline
&~~$k_1$~~&~~~~~~~~~~~$e_1^T$~~~~~~~~~~~\\
\hline
$e_2^L$&~$\displaystyle-2\sqrt{\al^\prime}\mathrm{k}_2E-\frac{iv}{2}$~&$0$\\
\hline
$e_2^T$&$0$&$1$\\
\hline
\end{tabular}}
~~
\subfigure[The first and the third particles]{
\begin{tabular}{|c|c|c|}
\hline
&$k_1$&~~~~~~~~~~~~~$e_1^T$~~~~~~~~~~~~~\\
\hline
$k_3$&$\ast$&$-\mathrm{k}_3\sin\phi$\\
\hline
$e_3^T$&~~$\displaystyle\frac{2\mathrm{k}_2\mathrm{k}_3\sin\phi+i\mathrm{k}_3V_0\sin\phi}{\sqrt{4\mathrm{k}_3^2+V_0^2}}$~~&~~$\displaystyle\frac{2\mathrm{k}_3\cos\phi+iV_0}{\sqrt{4\mathrm{k}_3^2+V_0^2}
}$~~\\
\hline
\end{tabular}}
~\\
~\\
\subfigure[The second and the third particles]{
\begin{tabular}{|c|c|c|c|}
  \hline
   & $k_2$ & $e_2^L$ & ~~~~~~$e_2^T$~~~~~~~ \\
  \hline
  $k_3$ & $\ast$ &~~ $\displaystyle\sqrt{\al^\prime}\big(\mathrm{k}_2E_3-\mathrm{k}_3E_2\cos\phi\big)+\frac{iv}{2}$ ~~& ~~ $-\mathrm{k}_3\sin\phi$~~ \\
  \hline
  $e_3^T$ & $\displaystyle-\frac{2\mathrm{k}_2\mathrm{k}_3\sin\phi+i\mathrm{k}_3V_0\sin\phi}{\sqrt{4\mathrm{k}_3^2+V_0^2}}$ & ~~$\displaystyle-\frac{2\sqrt{\al^\prime}\mathrm{k}_3E_2\sin\phi}{\sqrt{4\mathrm{k}_3^2+V_0^2}}$~~ & ~~$\displaystyle\frac{2\mathrm{k}_3\cos\phi+iV_0}{\sqrt{4\mathrm{k}_3^2+V_0^2}}$~~\\
  \hline
  \end{tabular}}
  \end{center}
  \caption{Momentum-polarization contractions between particles}\label{1}
  \end{table} 

%% file: IV-B-1.tex
Below we list the results for three-point functions in the light-like linear dilaton background. These seven sample processes all include photon ($P(\zeta, k)$) as the final outgoing particle and we list only one particular channel of the stringy scattering amplitudes.\\
~\\
\textbf{\footnotesize P($\zeta,k_1$)-T($k_2$)-T($k_3$)}
\beqa
\cA_{PTT}=\sqrt{2\al^\prime}\zeta^*\cdot k_2.
\eeqa
\textbf{\footnotesize P($\zeta_1,k_1$)-T($k_2$)-P($\zeta_3,k_3$)}
\beqa
\cA_{PTP}=-2\al^\prime\zeta^*_1\cdot k_2 \zeta_3\cdot k_2+\zeta_1^*\cdot\zeta_3.
\eeqa
\textbf{\footnotesize P($\zeta_1,k_1$)-P($\zeta_2,k_2$)-T($k_3$)}
\beqa
\cA_{PPT}=2\al^\prime\zeta^*_1\cdot k_2 \zeta_2\cdot k_3+\zeta_1^*\cdot\zeta_2.
\eeqa
\textbf{\footnotesize P($\zeta_1,k_1$)-P($\zeta_2,k_2$)-P($\zeta_3,k_3$)}
\beqa
\cA_{PPP}=-\sqrt{2\al^\prime}\zeta^*_1\cdot\zeta_2 \zeta_3\cdot k_2+\sqrt{2\al^\prime}\zeta^*_1\cdot\zeta_3 \zeta_2\cdot k_3+
\sqrt{2\al^\prime}\zeta_2\cdot\zeta_3 \zeta_1^*\cdot k_2.
\eeqa
\textbf{\footnotesize P($\zeta,k_1$)-M($\epsilon_{\mu\nu},k_2$)-T($k_3$)}
\beqa
\cA_{PMT}=\sqrt{2\al^\prime}\epsilon_{\mu\nu}\zeta^{*\mu}(k^{*\nu}_1+k^\nu_3)+\epsilon\cdot\zeta^*
   +(2\al^\prime)^{\frac{3}{2}}\zeta^*\cdot k_2\epsilon_{\mu\nu}k_1^{*\mu} k_3^\nu.
\eeqa
\textbf{\footnotesize P($\zeta_1,k_1$)-M($\epsilon_{\mu\nu},k_2$)-P($\zeta_3,k_3$)}
\beqa
\cA_{PMP}&=&2\epsilon_{\mu\nu}\zeta_1^{*\mu}\zeta_3^\nu-2\al^\prime\epsilon_{\mu\nu}\zeta_1^{*\mu} (k_1^{*\nu}+k_3^\nu) k_2\cdot \zeta_3
 +2\al^\prime\epsilon_{\mu\nu}\zeta_3^\mu (k_1^{*\nu}+k_3^\nu) k_2\cdot \zeta_1^*\\
 \notag&&+2\al^\prime\epsilon_{\mu\nu}k_1^{*\mu} k_3^\nu(\zeta_1^*\cdot\zeta_3-2\al^\prime\zeta_1^*\cdot k_2 \zeta_3 \cdot k_2)-\sqrt{2\al^\prime}\epsilon\cdot\zeta_1^* k_2\cdot\zeta_3-\sqrt{2\al^\prime}\epsilon\cdot\zeta_3 k_2\cdot\zeta_1^*.
 \eeqa
\textbf{\footnotesize P($\zeta,k_1$)-M($\epsilon^{(2)}_{\mu\nu},k_2$)-M($\epsilon^{(3)}_{\mu\nu},k_3$)}
\beqa
\notag&&\cA_{PMM}\\
\notag&=&\Big[\sqrt{2\al^\prime}\zeta^*\cdot k_2 \big(2\al^\prime \epsilon^{(3)}_{\rho\si}k^\rho_2k^\si_2-\sqrt{2\al^\prime}\epsilon^{(3)}\cdot k_2\big)
-2\sqrt{2\al^\prime}\epsilon^{(3)}_{\rho\si}\zeta^{*\rho}k_2^\si\Big]\big(2\al^\prime \epsilon^{(2)}_{\mu\nu}k^\mu_3 k^\nu_3-\sqrt{2\al^\prime}\epsilon^{(2)}\cdot k_3\big)\\
\notag&&+\sqrt{2\al^\prime}\zeta^{*}\cdot k_2\big(2\epsilon^{(2)}_{\mu\nu}\epsilon^{(3)\mu\nu}+4\sqrt{2\al^\prime}
\epsilon^{(3)}_{\rho\si}\epsilon^{(2)\rho}k^\si_2+4\sqrt{2\al^\prime}
\epsilon^{(2)}_{\rho\si}\epsilon^{(3)\rho}k^\si_3-8\al^\prime \epsilon^{(2)}_{\mu\nu}\epsilon^{(3)\nu\si}k_3^\mu k_{2\si}-6\epsilon^{(2)}\cdot\epsilon^{(3)}\big)\\
\notag&&+2\sqrt{2\al^\prime}\epsilon^{(2)}_{\mu\nu}\zeta^{*\mu} k^\nu_3\big(2\al^\prime \epsilon^{(3)}_{\rho\si}k^\rho_2k^\si_2-\sqrt{2\al^\prime}\epsilon^{(3)}\cdot k_2\big)
+4\epsilon^{(2)}_{\mu\nu}\zeta^{*\mu}\epsilon^{(3)\nu}-4\epsilon^{(3)}_{\mu\nu}\zeta^{*\mu}\epsilon^{(2)\nu}\\
&&+4\sqrt{2\al^\prime}\epsilon^{(2)}_{\mu\nu}\epsilon^{(3)\nu}_{\si}k_3^\mu\zeta^{*\si}-4\sqrt{2\al^\prime}\epsilon^{(2)}_{\mu\nu}\epsilon^{(3)\nu}_{\si}k_2^\si\zeta^{*\mu}.
\eeqa

%% file: IV-B-2.tex
In the section, based on the kinematic setup in section IV-A-1, we rewrite all three-point functions in the helicity representation. The three-point functions involving photons and tachyons are easy to compute, since $\zeta_i\propto e^T$.\\
~\\
\textbf{\footnotesize P($\zeta,k_1$)-T($k_2$)-T($k_3$)}
\beqa
\cA_{PTT}=\sqrt{2\al^\prime}\zeta^*\cdot k_2=0.
\eeqa
\textbf{\footnotesize P($\zeta_1,k_1$)-T($k_2$)-P($\zeta_3,k_3$)}
\beqa
\cA_{PTP}=-2\al^\prime\zeta^*_1\cdot k_2 \zeta_3\cdot k_2+\zeta_1^*\cdot\zeta_3=1.
\eeqa
\textbf{\footnotesize P($\zeta_1,k_1$)-P($\zeta_2,k_2$)-T($k_3$)}
\beqa
\cA_{PPT}=2\al^\prime\zeta^*_1\cdot k_2 \zeta_2\cdot k_3+\zeta_1^*\cdot\zeta_2=1.
\eeqa
\textbf{\footnotesize P($\zeta_1,k_1$)-P($\zeta_2,k_2$)-P($\zeta_3,k_3$)}
\beqa
\cA_{PPP}=-\sqrt{2\al^\prime}\zeta^*_1\cdot\zeta_2 \zeta_3\cdot k_2+\sqrt{2\al^\prime}\zeta^*_1\cdot\zeta_3 \zeta_2\cdot k_3+
\sqrt{2\al^\prime}\zeta_2\cdot\zeta_3 \zeta_1^*\cdot k_2=0.
\eeqa
For all the stringy scattering amplitudes involving the spin-two particle (M), we need the momentum decomposition, Eqs.\eqref{dec1}, \eqref{dec2} and \eqref{dec3}.\\
\textbf{\footnotesize P($\zeta,k_1$)-M($\epsilon_{\mu\nu},k_2$)-T($k_3$)}
\beqas
 \displaystyle a_4=1, \quad \displaystyle b_4=-1, \quad
\displaystyle a_5=-\frac{iv}{2},   \quad \displaystyle b_5=-1+\frac{iv}{2}.
\eeqas
\beqa
\notag\cA_{PMT}&=&\sqrt{2\al^\prime}\epsilon_{\mu\nu}\zeta^{*\mu}(k^{*\nu}_1+k^\nu_3)+\epsilon\cdot\zeta^*
   +(2\al^\prime)^{\frac{3}{2}}\underbrace{\zeta^*\cdot k_2}_{=0}\epsilon_{\mu\nu}k_1^{*\mu} k_3^\nu\\
\notag&=&\sqrt{2}\epsilon_{\mu\nu}\big(e^{T}\big)^{\mu}\bigg[\big(a_4+a_5\big)\big(e^P\big)^\nu
+\big(b_4+b_5\big)\big(e^L\big)^\nu\bigg]+\epsilon_\mu \big(e^{T}\big)^{\mu}\\
\notag&=&\sqrt{2}\bigg[\big(a_4+a_5\big)u^{PT}
+\big(b_4+b_5\big)u^{LT}\bigg]
-\sqrt 2 a_2u^{PT}-\sqrt 2b_2u^{LT}\\
&=&\sqrt{2}\bigg[(-iv)u^{PT}
+\big(-2+iv\big)u^{LT}\bigg].
\eeqa
\textbf{\footnotesize P($\zeta_1,k_1$)-M($\epsilon_{\mu\nu},k_2$)-P($\zeta_3,k_3$)}
\beqas
 \displaystyle a_4=\frac{1}{2}, \quad \displaystyle b_4=-\frac{1}{2},\quad\displaystyle a_5=-\frac{1}{2}-\frac{iv}{2}, \quad  \displaystyle b_5=-\frac{1}{2}+\frac{iv}{2}.
\eeqas
\beqa
\notag \cA_{PMP}&=&2\epsilon                _{\mu\nu}\zeta_1^{\ast\mu}\zeta_3^\nu-2\al^\prime\epsilon_{\mu\nu}\zeta_1^{\ast\mu} (k_1^{\ast\nu}+k_3^\nu) k_2\cdot \zeta_3
 +2\al^\prime\epsilon_{\mu\nu}\zeta_3^\mu (k_1^{\ast\nu}+k_3^\nu) k_2\cdot \zeta_1^\ast\\
\notag &&+2\al^\prime\epsilon_{\mu\nu}k_1^{\ast\mu} k_3^\nu(\zeta_1^\ast\cdot\zeta_3-2\al^\prime\zeta_1^\ast\cdot k_2 \zeta_3 \cdot k_2)-\sqrt{2\al^\prime}\epsilon\cdot\zeta_1^\ast k_2\cdot\zeta_3-\sqrt{2\al^\prime}\epsilon\cdot\zeta_3 k_2\cdot\zeta_1^\ast\\
 \notag&=& 2\epsilon_{\mu\nu}\zeta_1^{\ast\mu}\zeta_3^\nu+2\al^\prime\epsilon_{\mu\nu}k_1^{\ast\mu} k_3^\nu\\
 \notag&=& 2u^{TT}+2 \Big[a_4a_5u^{PP}+(a_4b_5+a_5b_4)u^{PL}+b_4b_5u^{LL}\Big]\\
 &=& 2u^{TT}-\big(\frac{1+iv}{2}\big)u^{PP}+\big(iv\big)u^{PL}+\big(\frac{1-iv}{2}\big)u^{LL}.
 \eeqa
\textbf{\footnotesize P($\zeta,k_1$)-M($\epsilon^{(2)}_{\mu\nu},k_2$)-M($\epsilon^{(3)}_{\mu\nu},k_3$)}
\beqa
\notag&&\hspace{3.5cm} \displaystyle a_4=0, \quad \displaystyle b_4=0, \quad
\displaystyle a_5=-1-\frac{iv}{2},  \quad  \displaystyle b_5=\frac{iv}{2}.\\
\notag&&\cA_{PMM}\\
\notag&=&\Big[\sqrt{2\al^\prime}\zeta^*\cdot k_2 \big(2\al^\prime \epsilon^{(3)}_{\rho\si}k^\rho_2k^\si_2-\sqrt{2\al^\prime}\epsilon^{(3)}\cdot k_2\big)
-2\sqrt{2\al^\prime}\epsilon^{(3)}_{\rho\si}\zeta^{*\rho}k_2^\si\Big]\big(2\al^\prime \epsilon^{(2)}_{\mu\nu}k^\mu_3 k^\nu_3-\sqrt{2\al^\prime}\epsilon^{(2)}\cdot k_3\big)\\
\notag&&+\sqrt{2\al^\prime}\zeta^{*}\cdot k_2\big(2\epsilon^{(2)}_{\mu\nu}\epsilon^{(3)\mu\nu}+4\sqrt{2\al^\prime}
\epsilon^{(3)}_{\rho\si}\epsilon^{(2)\rho}k^\si_2+4\sqrt{2\al^\prime}
\epsilon^{(2)}_{\rho\si}\epsilon^{(3)\rho}k^\si_3-8\al^\prime \epsilon^{(2)}_{\mu\nu}\epsilon^{(3)\nu\si}k_3^\mu k_{2\si}-6\epsilon^{(2)}\cdot\epsilon^{(3)}\big)\\
\notag&&+2\sqrt{2\al^\prime}\epsilon^{(2)}_{\mu\nu}\zeta^{*\mu} k^\nu_3\big(2\al^\prime \epsilon^{(3)}_{\rho\si}k^\rho_2k^\si_2-\sqrt{2\al^\prime}\epsilon^{(3)}\cdot k_2\big)
+4\epsilon^{(2)}_{\mu\nu}\zeta^{*\mu}\epsilon^{(3)\nu}-4\epsilon^{(3)}_{\mu\nu}\zeta^{*\mu}\epsilon^{(2)\nu}\\
\notag&&+4\sqrt{2\al^\prime}\epsilon^{(2)}_{\mu\nu}\epsilon^{(3)\nu}_{\si}k_3^\mu\zeta^{*\si}-4\sqrt{2\al^\prime}\epsilon^{(2)}_{\mu\nu}\epsilon^{(3)\nu}_{\si}k_2^\si\zeta^{*\mu}\\
\notag&=&-2\sqrt{2\al^\prime}\epsilon^{(3)}_{\rho\si}\zeta^{*\rho}k_2^\si\big(2\al^\prime \epsilon^{(2)}_{\mu\nu}k^\mu_3 k^\nu_3-\sqrt{2\al^\prime}\epsilon^{(2)}\cdot k_3\big)+2\sqrt{2\al^\prime}\epsilon^{(2)}_{\mu\nu}\zeta^{*\mu} k^\nu_3\big(2\al^\prime \epsilon^{(3)}_{\rho\si}k^\rho_2k^\si_2
-\sqrt{2\al^\prime}\epsilon^{(3)}\cdot k_2\big)\\
\notag&&+4\epsilon^{(2)}_{\mu\nu}\zeta^{*\mu}\big(-\sqrt{2\al^\prime}\epsilon^{(3)\nu\si}k_{2\si}+\epsilon^{(3)\nu}\big)
-4\epsilon^{(3)}_{\mu\nu}\zeta^{*\mu}\big(-\sqrt{2\al^\prime}\epsilon^{(2)\nu\si}k_{3\si}+\epsilon^{(2)\nu}\big)\\
&=&0.\eeqa
In the last case, we have $k_1=0$, see Eq.\eqref{dec1}. Thus, combining momentum conservation, $k_2+k_3+iV=0$, and the Virasoro constraint, Eq.\eqref{L1C}, we show that the three-point on-shell stringy scattering amplitude $\cA_{PMM}$ vanishes identically. 

%% file: IV-B-3.tex
We emphasize that in the calculations of all three-point functions, the full vertex operators, Eqs.\eqref{vetx1}, \eqref{vetx2}, \eqref{vetx3} are used for each physical state. Thus, using Eqs.\eqref{ZNSI1}, \eqref{ZNSI2}, \eqref{ZNSII2}, we can easily check that all massive stringy Ward identities \cite{Lee:1994wp} hold true at arbitrary value of $V_0$. These identities clearly demonstrate a smooth deformation of the target space gauge symmetry. In addition, we can examine the evolution of three-point coupling constants as a function of the moduli parameter, $V_0$. If we identify $V_0=0$ and $V_0/E\rightarrow \infty$ as two fixed-points in the spectral flow, then we can summarize the symmetry pattern of three-point functions in Table \ref{2}.
\begin{table}
\begin{center}
\begin{tabular}{|c|c|l|}
\hline
&~\textbf{Flat Background}~&~~~~~~~\textbf{Linear Dilaton Background}~~~~~~~  \\
\cline{2-3}
~\textbf{Process}~& $~~V_0=0~~$~~
&~~~~~~Finite $V_0$~~~~~~~~\hspace{-0.1pt}\vline
~~~~~$~~~~V_0/E\rightarrow\infty$\\
\hline
\footnotesize PTT&0&
~~~~~~~~~~~~0~~~~~~~~~~~~\hspace{1pt}\vline
~~~~~~~~~~~~~0~~~~~~~~\\
\hline
\footnotesize PTP&1&
~~~~~~~~~~~~1~~~~~~~~~~~~\hspace{1pt}\vline
~~~~~~~~~~~~~1~~~~~~~~\\
\hline
\footnotesize PPP&0&
~~~~~~~~~~~~0~~~~~~~~~~~~\hspace{1pt}\vline
~~~~~~~~~~~~~0~~~~~~~~\\
\hline
\footnotesize PM(LL)T&0&
~~~~~~~~~~~~0~~~~~~~~~~~~\hspace{1pt}\vline
~~~~~~~~~~~~~0~~~~~~~~\\
\hline
\footnotesize PM(LT)T &$-\sqrt 2$&
~~~$\displaystyle
\frac{-2\sqrt 2 +\sqrt 2iv}{\sqrt{v^2+4}}$~~~~\vline
~~~~~~~~~~~$\sqrt 2i$~~~~~~\\
\hline
\footnotesize PM(TT)T &0&
~~~~~~~~~~~~0~~~~~~~~~~~~\hspace{1pt}\vline
~~~~~~~~~~~~~0~~~~~~~~\\
\hline
\footnotesize PM(LL)P&$\displaystyle-\frac{3}{4}$&
$\displaystyle-\frac{4v^2+iv+3}{2\sqrt{8v^4+9v^2+4}}~\hspace{0.7pt}$\vline~~$~~~~~\displaystyle-\frac{1}{\sqrt 2}$\\
\hline
\footnotesize PM(LT)P&0&
~~~~~~~~~~~~0~~~~~~~~~~~~\hspace{1pt}\vline
~~~~~~~~~~~~~0~~~~~~~~\\
\hline
\footnotesize PM(TT)P&$\sqrt 2$&
~~~~~~~~~~$\sqrt 2$~~~~~~~~~~~~\hspace{-0.8pt}\vline
~~~~~~~~~~~$\sqrt 2$~~~~~~~~\\
\hline
\end{tabular}
\end{center}
\caption{Symmetry pattern of three-point coupling constants as functions of $V_0$}\label{2}
\end{table} 

%% file: IV-C-1.tex
    In the discussion of section IV-A, we have shown that, for our choice of kinematics, the first particle must be a massless photon. To extract the high-energy stringy symmetry, we list five sample stringy scattering amplitudes (s-t channel only) in the following:
~\\
~\\
\textbf{\footnotesize T($k_4$)-T($k_3$)-T($k_2$)-P($\zeta,k_1$)}\\
  \beqa
\notag\cA_{TTTP}=\frac{\Ga(-\al^\prime t-1)\Ga(-\al^\prime s-1)}{\Ga(-\al^\prime t-\al^\prime s-1)}\Big[(\al^\prime t+\al^\prime s+2)(\sqrt{2\al^\prime}\zeta\cdot k_2)+(\al^\prime s+1)(\sqrt{2\al^\prime}\zeta\cdot k_3)\Big].\\
\eeqa
\textbf{\footnotesize T($k_4$)-T($k_3$)-P($\zeta_2,k_2$)-P($\zeta_1,k_1$)}\\
\beqa
\notag&&\cA_{TTPP}=\frac{\Ga(-\al^\prime t-1)\Ga(-\al^\prime s-1)}{\Ga(-\al^\prime t-\al^\prime s)}
\left[\begin{array}{rl}
  (\al^\prime t+1)(\al^\prime t+\al^\prime s+1)(\zeta_1\cdot\zeta_2-&2\al^\prime \zeta_1\cdot k_2 \zeta_2\cdot k_1)\\
   -(\al^\prime t+1)(\al^\prime s+1)&(2\al^\prime\zeta_1\cdot k_3 \zeta_2\cdot k_1)\\+(\al^\prime t+\al^\prime s+1)(\al^\prime s+1)&(2\al^\prime\zeta_1\cdot k_2 \zeta_2\cdot k_3)\\
   +(\al^\prime s+1)(\al^\prime s)&(2\al^\prime\zeta_1\cdot k_3 \zeta_2\cdot k_3)
\end{array}\right].\\
\eeqa
\textbf{\footnotesize T($k_4$)-P($\zeta_3,k_3$)-P($\zeta_2,k_2$)-P($\zeta_1,k_1$)}\\
  \beqa
\notag&&\cA_{TPPP}
=\frac{\Ga(-\al^\prime t-1)\Ga(-\al^\prime s-1)}{\Ga(-\al^\prime t-\al^\prime s+1)}\times\\
\notag&&\hspace{-1.5cm}\times\left\{\begin{array}{r}
                 (\al^\prime t+\al^\prime s)\left\{
                 \begin{array}{rll}
                   (\al^\prime t+1)(\al^\prime t)&\Big[&(2\al^\prime)^{\frac{3}{2}}\zeta_1\cdot k_2\zeta_2\cdot k_1\zeta_3\cdot k_1-\sqrt{2\al^\prime}\zeta_3\cdot k_1\zeta_1\cdot\zeta_2\Big]\\
                   +(\al^\prime t+1)(\al^\prime s+1)&\Big[-& (2\al^\prime)^{\frac{3}{2}}\zeta_1\cdot k_2\zeta_2\cdot k_3\zeta_3\cdot k_1
+(2\al^\prime)^{\frac{3}{2}}\zeta_1\cdot k_3\zeta_2\cdot k_1\zeta_3\cdot k_2\Big]\\
                   +(\al^\prime s+1)(\al^\prime s)&\Big[-& (2\al^\prime)^{\frac{3}{2}}\zeta_1\cdot k_3\zeta_2\cdot k_3\zeta_3\cdot k_2+\sqrt{2\al^\prime}\zeta_1\cdot k_3\zeta_2\cdot\zeta_3\big)\Big]
                 \end{array}\right\}
                 \\
                 +(\al^\prime t+1)(\al^\prime s+1)\left\{\begin{array}{rll}
                 (\al^\prime t)&\Big[&\hspace{-2mm}(2\al^\prime)^{\frac{3}{2}}\zeta_1\cdot k_3\zeta_2\cdot k_1\zeta_3\cdot k_1-\sqrt{2\al^\prime}\zeta_2\cdot k_1\zeta_1\cdot\zeta_3\Big]\\
                 +(\al^\prime s)&\Big[-&\hspace{-2mm}(2\al^\prime)^{\frac{3}{2}}\zeta_1\cdot k_3\zeta_2\cdot k_3\zeta_3\cdot k_1+\sqrt{2\al^\prime}\zeta_2\cdot k_3\zeta_1\cdot\zeta_3\Big]
                 \end{array}\right\}
                  \\
                  +(\al^\prime t+\al^\prime s+1)(\al^\prime t+\al^\prime s)\left\{
                  \begin{array}{rll}
                  (\al^\prime t+1)&\Big[&(2\al^\prime)^{\frac{3}{2}}\zeta_1\cdot k_2\zeta_2\cdot k_1\zeta_3\cdot k_2-\sqrt{2\al^\prime}\zeta_3\cdot k_2\zeta_1\cdot\zeta_2\Big]\\
                  +(\al^\prime s+1)&\Big[-&(2\al^\prime)^{\frac{3}{2}}\zeta_1\cdot k_2\zeta_2\cdot k_3\zeta_3\cdot k_2+\sqrt{2\al^\prime}\zeta_1\cdot k_2\zeta_2\cdot\zeta_3\Big]
                  \end{array}
                  \right\}
               \end{array}\right\}.\\
\eeqa
\textbf{\footnotesize T($k_4$)-T($k_3$)-M($\epsilon_{\mu\nu},k_2$)-P($\zeta,k_1$)}\\
  \beqa
\notag&&\cA_{TTMP}
=\frac{\Ga(-\al^\prime t-1)\Ga(-\al^\prime s-1)}{\Ga(-\al^\prime t-\al^\prime s+1)}\times\\
\notag&&\times\left\{\begin{array}{r}
(\al^\prime t+\al^\prime s)\left\{\begin{array}{rl}(\al^\prime t+1)(\al^\prime t)
&\left[\begin{array}{l}~~\sqrt{2\al^\prime}\zeta\cdot k_2\big(2\al^\prime\epsilon_{\mu\nu}k^\mu_1k^\nu_1-\sqrt{2\al^\prime}\epsilon\cdot k_1\big)\\
-2\sqrt{2\al^\prime}\epsilon_{\mu\nu}k_1^\mu\zeta^\nu+2\epsilon\cdot\zeta\end{array}\right]\\
+(\al^\prime t+1)(\al^\prime s+1)&\Big[-2(2\al^\prime)^{\frac{3}{2}}\zeta\cdot k_2\epsilon_{\mu\nu}k_1^\mu k_3^\nu+2\sqrt{2\al^\prime}\epsilon_{\mu\nu}\zeta^\mu k_3^\nu\Big]\\
+(\al^\prime s+1)(\al^\prime s)&\Big[~~\sqrt{2\al^\prime}\zeta\cdot k_2\big(2\al^\prime\epsilon_{\mu\nu}k^\mu_3k^\nu_3-\sqrt{2\al^\prime}\epsilon\cdot k_3\big)\Big]\end{array}\right\}\\
+(\al^\prime s+1)\left\{\begin{array}{rll}(\al^\prime t+1)(\al^\prime t)&\Big[&\sqrt{2\al^\prime}\zeta\cdot k_3\big(2\al^\prime\epsilon_{\mu\nu}k^\mu_1k^\nu_1-\sqrt{2\al^\prime}\epsilon\cdot k_1\big)\Big]\\
+(\al^\prime t+1)(\al^\prime s)&\Big[-&2(2\al^\prime)^{\frac{3}{2}}\zeta\cdot k_3\epsilon_{\mu\nu}k_1^\mu k_3^\nu\Big]\\
+(\al^\prime s)(\al^\prime s-1)&\Big[&\sqrt{2\al^\prime}\zeta\cdot k_3\big(2\al^\prime\epsilon_{\mu\nu}k^\mu_3k^\nu_3-\sqrt{2\al^\prime}\epsilon\cdot k_3\big)\Big]\end{array}\right\}
\end{array}\right\}.\\
\eeqa
\textbf{\footnotesize T($k_4$)-P($\zeta_3,k_3$)-M($\epsilon_{\mu\nu},k_2$)-P($\zeta_1,k_1$)}\\
\beqa
&&\cA_{TPMP}=\frac{\Ga(-\al^\prime t-1)\Ga(-\al^\prime s-1)}{\Ga(-\al^\prime t-\al^\prime s+2)}\times\\
\notag&&\hspace{-2.3cm}\times\left\{\begin{array}{r}
(\al^\prime t+1)(\al^\prime s+1)
\left\{\begin{array}{rl}(\al^\prime t)(\al^\prime t-1)&\Big[\big(\zeta_1\cdot\zeta_3-2\al^\prime\zeta_1\cdot k_3 \zeta_3\cdot k_1\big)\big(2\al^\prime\epsilon_{\mu\nu}k^\mu_1k^\nu_1-\sqrt{2\al^\prime}\epsilon\cdot k_1\big)\Big]\\
+(\al^\prime t)(\al^\prime s)&\Big[\big(\zeta_1\cdot\zeta_3-2\al^\prime\zeta_1\cdot k_3 \zeta_3\cdot k_1\big)\big(-4\al^\prime \epsilon_{\mu\nu}k_1^\mu k_3^\nu\big)\Big]\\
+(\al^\prime s)(\al^\prime s-1)&\Big[\big(\zeta_1\cdot\zeta_3-2\al^\prime\zeta_1\cdot k_3 \zeta_3\cdot k_1\big)\big(2\al^\prime\epsilon_{\mu\nu}k^\mu_3k^\nu_3-\sqrt{2\al^\prime}\epsilon\cdot k_3\big)\Big]
\end{array}\right\}\\
+(\al^\prime t+\al^\prime s-1)\left\{\begin{array}{rl}
(\al^\prime t+1)(\al^\prime t)(\al^\prime t-1)&\left[\begin{array}{r}
-2\al^\prime\zeta_1\cdot k_2\zeta_3\cdot k_1\big(2\al^\prime\epsilon_{\mu\nu}k^\mu_1k^\nu_1-\sqrt{2\al^\prime}\epsilon\cdot k_1\big)\\
+2\sqrt{2\al^\prime} \zeta_3\cdot k_1\big(\sqrt{2\al^\prime}\epsilon_{\mu\nu}k^\mu_1\zeta_1^\nu-\epsilon\cdot\zeta_1\big)\end{array}\right]\\
+(\al^\prime t+1)(\al^\prime t)(\al^\prime s+1)&\left[\begin{array}{r}-2\al^\prime\zeta_1\cdot k_3\zeta_3\cdot k_2\big(2\al^\prime\epsilon_{\mu\nu}k^\mu_1k^\nu_1-\sqrt{2\al^\prime}\epsilon\cdot k_1\big)\\
+2(2\al^\prime)^2\zeta_1\cdot k_2 \zeta_3\cdot k_1\epsilon_{\mu\nu}k^\mu_1k^\nu_3-4\al^\prime\zeta_3\cdot k_1\epsilon_{\mu\nu}\zeta^\mu_1k^\nu_3\end{array}\right]\\
+(\al^\prime t+1)(\al^\prime s+1)(\al^\prime s)&\left[\begin{array}{r}-2\al^\prime\zeta_1\cdot k_2\zeta_3\cdot k_1\big(2\al^\prime\epsilon_{\mu\nu}k^\mu_3k^\nu_3-\sqrt{2\al^\prime}\epsilon\cdot k_3\big)
\\+2(2\al^\prime)^2\zeta_1\cdot k_3 \zeta_3\cdot k_2\epsilon_{\mu\nu}k^\mu_1k^\nu_3-4\al^\prime\zeta_1\cdot k_3\epsilon_{\mu\nu}k^\mu_1\zeta^\nu_3\end{array}\right]\\
+(\al^\prime s+1)(\al^\prime s)(\al^\prime s-1)&\left[\begin{array}{r}-2\al^\prime\zeta_1\cdot k_3\zeta_3\cdot k_2\big(2\al^\prime\epsilon_{\mu\nu}k^\mu_3k^\nu_3-\sqrt{2\al^\prime}\epsilon\cdot k_3\big)
\\+2\sqrt{2\al^\prime} \zeta_1\cdot k_3\big(\sqrt{2\al^\prime}\epsilon_{\mu\nu}k^\mu_3\zeta_3^\nu-\epsilon\cdot\zeta_3\big)\end{array}\right]
\end{array}\right\}\\
+(\al^\prime t+\al^\prime s)(\al^\prime t+\al^\prime s-1)\left\{\begin{array}{rl}
(\al^\prime t+1)(\al^\prime t)&\left[\begin{array}{r}-2\al^\prime\zeta_1\cdot k_2\zeta_3\cdot k_2\big(2\al^\prime\epsilon_{\mu\nu}k^\mu_1k^\nu_1-\sqrt{2\al^\prime}\epsilon\cdot k_1\big)
\\+2\sqrt{2\al^\prime} \zeta_3\cdot k_2\big(\sqrt{2\al^\prime}\epsilon_{\mu\nu}k^\mu_1\zeta_1^\nu-\epsilon\cdot\zeta_1\big)\end{array}\right]\\
+(\al^\prime t+1)(\al^\prime s+1)&\left[\begin{array}{r}2(2\al^\prime)^2\zeta_1\cdot k_2\zeta_3\cdot k_2 \epsilon_{\mu\nu}k^\mu_1k^\nu_3-4\al^\prime \zeta_1\cdot k_2 \epsilon_{\mu\nu}k^\mu_1\zeta^\nu_3\\
+2\epsilon_{\mu\nu}\zeta_1^\mu\zeta_3^\nu-4\al^\prime \zeta_3\cdot k_2 \epsilon_{\mu\nu}\zeta^\mu_1 k^\nu_3\end{array}\right]\\
+(\al^\prime s+1)(\al^\prime s)&\left[\begin{array}{r}-2\al^\prime\zeta_1\cdot k_2\zeta_3\cdot k_2\big(2\al^\prime\epsilon_{\mu\nu}k^\mu_3k^\nu_3-\sqrt{2\al^\prime}\epsilon\cdot k_3\big)
\\+2\sqrt{2\al^\prime} \zeta_1\cdot k_2\big(\sqrt{2\al^\prime}\epsilon_{\mu\nu}k^\mu_3\zeta_3^\nu-\epsilon\cdot\zeta_3\big)\end{array}\right]
\end{array}\right\}\end{array}
\right\}.
\eeqa

%% file: IV-C-2.tex
\begin{center}\small\textsl{2-1. Kinematics}\end{center}
  \input{IV-C-2-1}
  \newpage
  \begin{center}\small\textsl{2-2. High-energy limits of stringy scattering amplitudes in the flat space-time background}\end{center}
  \input{IV-C-2-2}

%% file: IV-C-2-1.tex
\par In order to make comparison between the stringy symmetry at two fixed-points, we first study the high-energy limits of stringy scattering amplitudes in the flat space-time background. In this limit, the Mandelstam variables become
\beqa
s&=&4E^2,\\
\notag t&=&\big(E_2-E_3\big)^2-\big(\mathrm{k}_2-\mathrm{k}_3\cos\phi\big)^2-\mathrm{k}_3^2\sin^2\phi\\
&\sim&-4E^2\sin^2\frac{\phi}{2}+(m_1^2+m_2^2+m^2_3+m_4^2)\sin^2\frac{\phi}{2}.
\eeqa
Notice that it is important to keep subleading terms in the $\frac{1}{\al^\prime E^2}$ expansion in the calculations of high-energy limits of stringy scattering amplitudes in the flat space-time. The stringy scattering amplitudes involving $\al^L_{-1}$ oscillator in general have lower energy power than the expectations from the naive power counting \cite{Chan:2003ee,Chan:2004yz,Chan:2005ne,Chan:2005ji}.
\par The fixed-angle high-energy limits of relevant momentum-polarization contractions are collected in Table \ref{3}.
\begin{table}
\centering
\subtable[The first and the second particles]{
\begin{tabular}{|c|c|c|}
\hline
&$k_1$&~~~~~~~~~~~$e_1^T$~~~~~~~~~~~\\
\hline
$e_2^L$&$\displaystyle-\sqrt{\al^\prime}\big(2E^2-\al_{12}\big)$&$0$\\
\hline
$e_2^T$&$0$&$1$\\
\hline
\end{tabular}}
~~~
\subtable[The first and the third particles]{
\begin{tabular}{|c|c|c|}
\hline
&$k_1$&~~~~~~~~~~~$e_1^T$~~~~~~~~~~~\\
\hline
$k_3$&$\ast$&$\displaystyle\big(-E+\frac{\al_{34}}{2E}\big)\sin\phi$\\
\hline
$e_3^T$&$\displaystyle\big(E-\frac{\al_{12}}{2E}\big)\sin\phi$&$\displaystyle\cos\phi$\\
\hline
\end{tabular}}
~\\
~\\
\subtable[The second and the third particles]{
\hspace{-1cm}\begin{tabular}{|c|c|c|c|}
  \hline
   & $k_2$ & $e_2^L$ & ~~~~~~$e_2^T$~~~~~~~ \\
  \hline
  $k_3$ & $\ast$ & $\displaystyle2\sqrt{\al^\prime}E^2\sin^2\frac{\phi}{2}+\frac{\sqrt{\al^\prime}\big(-2\al_{12}+\be_{34}\big)}{4}+\frac{\sqrt{\al^\prime}\big(2\al_{34}+\be_{12}\big)\cos\phi}{4}$ & $\displaystyle\big(-E+\frac{\al_{34}}{2E}\big)\sin\phi$ \\
  \hline
  $e_3^T$ & $\displaystyle\big(-E+\frac{\al_{12}}{2E}\big)\sin\phi$ & $\displaystyle\sqrt{\al^\prime}\big(E-\frac{\be_{12}}{4E}\big)\sin\phi$ & $\displaystyle\cos\phi$\\
  \hline
  \end{tabular}}
  \caption{Momentum-polarization contractions ($V=0$ and $E\rightarrow \infty$)between particles}\label{3}
  \end{table} 

%% file: IV-C-2-2.tex
\par Here we set $V_0=0$ and list the fixed-angle high-energy limits of the five sample stringy scattering amplitudes.\\
~\\
\textbf{\footnotesize T($k_4$)-T($k_3$)-T($k_2$)-P($\zeta,k_1$)}
\beqa
  \cA_{TTTP}\sim\frac{\Ga(-\al^\prime t-1)\Ga(-\al^\prime s-1)}{\Ga(-\al^\prime t-\al^\prime s)}\Big[2^{\frac{11}{2}}\al^{\prime\frac{5}{2}}E^5\sin\frac{\phi}{2}\cos^3\frac{\phi}{2}\Big].
\eeqa
\textbf{\footnotesize T($k_4$)-T($k_3$)-P($\zeta_2,k_2$)-P($\zeta_1,k_1$)}
  \beqa
  \cA_{TTPP}\sim\frac{\Ga(-\al^\prime t-1)\Ga(-\al^\prime s-1)}{\Ga(-\al^\prime t-\al^\prime s)}\Big[2^{\frac{14}{2}}\al^{\prime\frac{6}{2}}E^6\sin^2\frac{\phi}{2}\cos^2\frac{\phi}{2}\Big].
\eeqa
\textbf{\footnotesize T($k_4$)-P($\zeta_3,k_3$)-P($\zeta_2,k_2$)-P($\zeta_1,k_1$)}
\beqa
 \cA_{TPPP}\sim\frac{\Ga(-\al^\prime t-1)\Ga(-\al^\prime s-1)}{\Ga(-\al^\prime t-\al^\prime s)}\Big[-2^{\frac{17}{2}}\al^{\prime\frac{7}{2}}E^7\sin^3\frac{\phi}{2}\cos\frac{\phi}{2}\Big].
 \eeqa
~\\
For spin-two particle, M($\epsilon_{\mu\nu},k_2$), there are three independent polarizations:\\
 ~\\
\textbf{\footnotesize T($k_4$)-T($k_3$)-M($\epsilon_{\mu\nu},k_2$)-P($\zeta,k_1$)}
\begin{enumerate}
    \item[] $\al^L_{-1}\al^L_{-1}$:
    \beqa
    \cA_{TTMP}(LL)\sim\frac{\Ga(-\al^\prime t-1)\Ga(-\al^\prime s-1)}{\Ga(-\al^\prime t-\al^\prime s)}\Big[2^{\frac{13}{2}}\al^{\prime\frac{7}{2}}E^7\sin^3\frac{\phi}{2}\cos\frac{\phi}{2}\Big]
    .\eeqa
    \item[]$\al^L_{-1}\al^{T}_{-1}$:
    \beqa
    \label{TTMPLT}\cA_{TTMP}(LT)\sim\frac{\Ga(-\al^\prime t-1)\Ga(-\al^\prime s-1)}{\Ga(-\al^\prime t-\al^\prime s)}\Big[2^{\frac{13}{2}}\al^{\prime\frac{6}{2}}E^6\sin^2\frac{\phi}{2}\cos\phi\Big]
    .\eeqa
    \item[]$\al^T_{-1}\al^T_{-1}$:
    \beqa
    \cA_{TTMP}(TT)\sim\frac{\Ga(-\al^\prime t-1)\Ga(-\al^\prime s-1)}{\Ga(-\al^\prime t-\al^\prime s)}\Big[2^{\frac{17}{2}}\al^{\prime\frac{7}{2}}E^7\sin^3\frac{\phi}{2}\cos\frac{\phi}{2}\Big].
    \eeqa
  \end{enumerate}
\textbf{\footnotesize T($k_4$)-P($\zeta_3,k_3$)-M($\epsilon_{\mu\nu},k_2$)-P($\zeta_1,k_1$)}
  \begin{enumerate}
   \item[]$\al^L_{-1}\al^L_{-1}$:
   \beqa
   \cA_{TPMP}(LL)\sim\frac{\Ga(-\al^\prime t-1)\Ga(-\al^\prime s-1)}{\Ga(-\al^\prime t-\al^\prime s)}\Big[-2^{\frac{16}{2}}\al^{\prime\frac{8}{2}}E^8\sin^4\frac{\phi}{2}\Big]
    .\eeqa
    \item[]$\al^L_{-1}\al^T_{-1}$:
    \beqa
    \label{TPMPLT}\cA_{TPMP}(LT)\sim\frac{\Ga(-\al^\prime t-1)\Ga(-\al^\prime s-1)}{\Ga(-\al^\prime t-\al^\prime s)}\Big[-2^{\frac{16}{2}}\al^{\prime\frac{7}{2}}E^7\sin^2\frac{\phi}{2}\tan\frac{\phi}{2}\cos\phi\Big]
    .\eeqa
    \item[]$\al^T_{-1}\al^T_{-1}$:
    \beqa
    \cA_{TPMP}(TT)\sim\frac{\Ga(-\al^\prime t-1)\Ga(-\al^\prime s-1)}{\Ga(-\al^\prime t-\al^\prime s)}\Big[-2^{\frac{20}{2}}\al^{\prime\frac{8}{2}}E^8\sin^4\frac{\phi}{2}\Big].
  \eeqa
  \end{enumerate}

%% file: IV-C-3.tex
\begin{center}\small\textsl{3-1. Kinematics}\end{center}
  \input{IV-C-3-1}
  \begin{center}\small\textsl{3-2. High-energy limits of stringy scattering amplitudes in the light-like linear dilaton background}\end{center}
  \input{IV-C-3-2}

%% file: IV-C-3-1.tex
\par In this subsection, we collect all relevant kinematic variables in the fixed-angle high-energy limit with infinite light-like linear dilaton gradient, $V_0/E\rightarrow\infty$.
\par First of all, the Mandelstam variables become
  \beqa
  &&s\sim 2iEV_0,\\
  &&t\sim-i\big(E_2-E_3\big)V_0-i\big(\mathrm{k}_2-\mathrm{k}_3\cos\phi\big)V_0\sim-2iEV_0\sin^2\frac{\phi}{2}.
  \eeqa
  It is interesting to see that, in the infinite light-like linear dilaton gradient limit $V_0/E\rightarrow\infty$, the ratio between two Mandelstam variables $t/s$ is the same as that of flat space-time background
  \beqas
  \bigg(\frac{t}{s}\bigg)\bigg|_{E\rightarrow\infty, V_0=0}=\bigg(\frac{t}{s}\bigg)\bigg|_{E\rightarrow\infty, \frac{V_0}{E}\rightarrow\infty}=-\sin^2\frac{\phi}{2}.
  \eeqas
 \par The relevant momentum-polarization contractions in the background with infinite light-like dilaton gradient are collected in Table \ref{4}.
\begin{table}[h]
\begin{center}
\subfigure[the first and
 the second particles]{
\begin{tabular}{|c|c|c|}
\hline
&$k_1$&~~~~~~~~~$e_1^T$~~~~~~~~~\\
\hline
$e_2^L$&~~~$\displaystyle-i\sqrt{\al^{\prime}}EV_0~~~$&$0$\\
\hline
$e_2^T$&$0$&$1$\\
\hline
\end{tabular}}
~~~
\subfigure[the first and
the third particles]{
\begin{tabular}{|c|c|c|}
\hline
&~~~$k_1~~~$&~~~~~~~~~$e_1^T$~~~~~~~~~\\
\hline
$k_3$&$\ast$&$-E\sin\phi$\\
\hline
$e_3^T$&~~~~~$\displaystyle iE\sin\phi$~~~~~&$i$\\
\hline
\end{tabular}}
~\\
~\\
\subfigure[the second and the third particles]{
\begin{tabular}{|c|c|c|c|}
  \hline
   & $k_2$ & $e_2^L$ &~~~~$e_2^T$~~~~~ \\
  \hline
  $k_3$ & $\ast$ & $~~~~~~\displaystyle i\sqrt{\al^{\prime}}EV_0$~~~~~~&~~~~ $\displaystyle-iE\sin\phi$ ~~~~~\\
  \hline
  $e_3^T$ & ~~~~$-iE\sin\phi$~~~~ & 0 & $i$\\
  \hline
  \end{tabular}}
  \end{center}
  \caption{Momentum-polarization contractions ($E\rightarrow \infty$ and $V_0/E\rightarrow \infty $) between particles}\label{4}
  \end{table} 

%% file: IV-C-3-2.tex
\par Here we take $V_0/E\rightarrow\infty$ limit, and list the fixed-angle high-energy limits of the five sample stringy scattering amplitudes.
~\\
~\\
\textbf{\footnotesize T($k_4$)-T($k_3$)-T($k_2$)-P($\zeta,k_1$)}
\beqa
\notag\cA_{TTTP}&\sim&\frac{\Ga(-\al^\prime t-1)\Ga(-\al^\prime s-1)}{\Ga(-\al^\prime t-\al^\prime s)}\big(-\al^\prime t-\al^\prime s\big)(\al^\prime s)(\sqrt{2\al^\prime}\zeta\cdot k_3)\\
&\sim&\frac{\Ga(-\al^\prime t-1)\Ga(-\al^\prime s-1)}{\Ga(-\al^\prime t-\al^\prime s)}\Big[-2^{\frac{7}{2}}\al^{\prime\frac{5}{2}}E^3V_0^2\sin\frac{\phi}{2}\cos^3\frac{\phi}{2} \Big]
.\eeqa
~\\
\textbf{\footnotesize T($k_4$)-T($k_3$)-P($\zeta_2,k_2$)-P($\zeta_1,k_1$)}\\
\beqa
\notag\cA_{TTPP}&\sim&\frac{\Ga(-\al^\prime t)\Ga(-\al^\prime s-1)}{\Ga(-\al^\prime t-\al^\prime s)}(\al^\prime s)^2(2\al^\prime\zeta_1\cdot k_3 \zeta_2\cdot k_3)\\
&\sim&\frac{\Ga(-\al^\prime t-1)\Ga(-\al^\prime s-1)}{\Ga(-\al^\prime t-\al^\prime s)}
\Big[-2^{\frac{10}{2}}\al^{\prime3}E^4V_0^2\sin^2\frac{\phi}{2}\cos^2\frac{\phi}{2}\Big]
.\eeqa
\\
\textbf{\footnotesize T($k_4$)-P($\zeta_3,k_3$)-P($\zeta_2,k_2$)-P($\zeta_1,k_1$)}\\
\beqa
\notag\cA_{TPPP}&\sim& \frac{\Ga(-\al^\prime t-1)\Ga(-\al^\prime s-1)}{\Ga(-\al^\prime t-\al^\prime s+1)}(\al^\prime s+1)(\al^\prime s)\times\\
\notag&&\times\left\{\begin{array}{r}
(\al^\prime t+1)
\Big[-(2\al^\prime)^{\frac{3}{2}}\zeta_1\cdot k_3\zeta_2\cdot k_3\zeta_3\cdot k_1\Big]\\
+(\al^\prime t+\al^\prime s)
\Big[-(2\al^\prime)^{\frac{3}{2}}\zeta_1\cdot k_3 \zeta_2\cdot k_3\zeta_3\cdot k_2\Big]
\end{array}\right\}\\
&\sim& \frac{\Ga(-\al^\prime t-1)\Ga(-\al^\prime s-1)}{\Ga(-\al^\prime t-\al^\prime s)}\bigg[2^{\frac{13}{2}}i\al^{\prime\frac{7}{2}}E^5V_0^2\sin^3\frac{\phi}{2}\cos\frac{\phi}{2}\bigg].
\eeqa
\textbf{\footnotesize T($k_4$)-T($k_3$)-M($\epsilon_{\mu\nu},k_2$)-P($\zeta,k_1$)}\\
\begin{description}
  \item[]$\al^L_{-1}\al^L_{-1}$:
  \beqa
\notag\cA_{TTMP}(LL)&\sim&\frac{\Ga(-\al^\prime t-1)\Ga(-\al^\prime s-1)}{\Ga(-\al^\prime t-\al^\prime s+1)}(\al^\prime s+1)\times\\
\notag&&\times\left\{\begin{array}{rll}(\al^\prime t+1)(\al^\prime t)&\Big[&(2\al^\prime)^{\frac{3}{2}}\zeta\cdot k_3\epsilon_{\mu\nu}k^\mu_1k^\nu_1\Big]\\
+(\al^\prime t+1)(\al^\prime s)&\Big[-2&(2\al^\prime)^{\frac{3}{2}}\zeta\cdot k_3\epsilon_{\mu\nu}k_1^\mu k_3^\nu\Big]\\
+(\al^\prime s)(\al^\prime s-1)&\Big[&(2\al^\prime)^{\frac{3}{2}}\zeta\cdot k_3\epsilon_{\mu\nu}k^\mu_3k^\nu_3\Big]
\end{array}\right\}\\
\notag&\sim& \frac{\Ga(-\al^\prime t-1)\Ga(-\al^\prime s-1)}{\Ga(-\al^\prime t-\al^\prime s+1)}(\al^\prime s)\Big[(\al^\prime t)^2+2(\al^\prime t)(\al^\prime s)+(\al^\prime s)^2\Big]\times\\
\notag&&\times\Big[(2\al^\prime)^{\frac{3}{2}}e^T_1\cdot k_3(e_2^L\cdot k_1)^2\Big]\\
&\sim&\frac{\Ga(-\al^\prime t-1)\Ga(-\al^\prime s-1)}{\Ga(-\al^\prime t-\al^\prime s)}\Big[2^{\frac{9}{2}}\al^{\prime\frac{9}{2}}E^5V_0^4\sin\frac{\phi}{2}\cos^3\frac{\phi}{2}\Big].
\eeqa
  \item[]$\al^L_{-1}\al^T_{-1}$:
  \beqa
\notag\cA_{TTMP}(LT)&\sim&\frac{\Ga(-\al^\prime t-1)\Ga(-\al^\prime s-1)}{\Ga(-\al^\prime t-\al^\prime s+1)}(-\al^\prime s-1)\times\\
\notag&&\times\left\{\begin{array}{rll}(\al^\prime t+1)(\al^\prime s)&\Big[2&(2\al^\prime)^{\frac{3}{2}}\zeta\cdot k_3\epsilon_{\mu\nu}k_1^\mu k_3^\nu\Big]\\
+(\al^\prime s)(\al^\prime s-1)&\Big[&(2\al^\prime)^{\frac{3}{2}}\zeta\cdot k_3\epsilon_{\mu\nu}k^\mu_3k^\nu_3)\Big]\end{array}\right\}
\\
\notag&\sim&\frac{\Ga(-\al^\prime t-1)\Ga(-\al^\prime s-1)}{\Ga(-\al^\prime t-\al^\prime s+1)}\times\\
\notag&&\times\bigg\{(\al^\prime s)^2\Big[-(\al^\prime t)(e^L_2\cdot k_1)+(\al^\prime s)(e^L_2\cdot k_3)\Big]\Big[(2\al^\prime)^{\frac{3}{2}}e^T_1\cdot k_3 e^T_2\cdot k_3\Big]\bigg\}\\
&\sim&\frac{\Ga(-\al^\prime t-1)\Ga(-\al^\prime s-1)}{\Ga(-\al^\prime t-\al^\prime s)}\Big[i2^{\frac{11}{2}}\al^{\prime\frac{8}{2}}E^5V_0^3\sin^2\frac{\phi}{2}\cos^2\frac{\phi}{2}\Big]
.\eeqa
\item[]$\al^T_{-1}\al^T_{-1}$:
  \beqa
\notag\cA_{TTMP}(TT)&\sim&\frac{\Ga(-\al^\prime t-1)\Ga(-\al^\prime s-1)}{\Ga(-\al^\prime t-\al^\prime s+1)}(\al^\prime s+1)(\al^\prime s)(\al^\prime s-1)\Big[(2\al^\prime)^{\frac{3}{2}}\zeta\cdot k_3\epsilon_{\mu\nu}k^\mu_3k^\nu_3\Big]\\
\notag&\sim&\frac{\Ga(-\al^\prime t-1)\Ga(-\al^\prime s-1)}{\Ga(-\al^\prime t-\al^\prime s)}\bigg[\frac{(\al^\prime s)^3(2\al^\prime)^{\frac{3}{2}}(e^T_1\cdot k_3)^3}{(-\al^\prime t-\al^\prime s)}\bigg]\\
&\sim&\frac{\Ga(-\al^\prime t-1)\Ga(-\al^\prime s-1)}{\Ga(-\al^\prime t-\al^\prime s)}\bigg[-2^{\frac{13}{2}}\al^{\prime\frac{7}{2}}E^5V_0^2\sin^3\frac{\phi}{2}\cos\frac{\phi}{2}\bigg]
.\eeqa
\item[]$\al^L_{-2}$:
\beqa
\notag\cA_{TTMP}(L)&\sim&\frac{\Ga(-\al^\prime t-1)\Ga(-\al^\prime s+1)}{\Ga(-\al^\prime t-\al^\prime s+1)}(\al^\prime s+1)\left[\begin{array}{r}(\al^\prime t+1)(\al^\prime t)(-2\al^\prime\zeta\cdot k_3e^L_2\cdot k_1)\\
+(\al^\prime s)(\al^\prime s-1)(-2\al^\prime\zeta\cdot k_3e^L_2\cdot k_3)\end{array}\right]\\
\notag&\sim&\frac{\Ga(-\al^\prime t-1)\Ga(-\al^\prime s-1)}{\Ga(-\al^\prime t-\al^\prime s)}\frac{(\al^\prime s)\Big[(\al^\prime t)^2-(\al^\prime s)^2\Big](-2\al^\prime e^T_1\cdot k_3e^L_2\cdot k_1)}{(-\al^\prime t-\al^\prime s)}\\
&\sim&\frac{\Ga(-\al^\prime t-1)\Ga(-\al^\prime s-1)}{\Ga(-\al^\prime t-\al^\prime s)}\Big[2^4i\al^{\prime\frac{7}{2}}E^4V_0^3\big(1+\sin^2\frac{\phi}{2}\big)\sin\phi\Big]
.\eeqa
\item[]$\al^T_{-2}$:
\beqa
\notag\cA_{TTMP}(T)&\sim&\frac{\Ga(-\al^\prime t-1)\Ga(-\al^\prime s-1)}{\Ga(-\al^\prime t-\al^\prime s+1)}(\al^\prime s+1)(\al^\prime s)(\al^\prime s-1)(-2\al^\prime\zeta\cdot k_3e^T_2\cdot k_3)\\
&\sim&\frac{\Ga(-\al^\prime t-1)\Ga(-\al^\prime s-1)}{\Ga(-\al^\prime t-\al^\prime s)}\Big[-2^5\al^{\prime 3}E^4V_0^2\sin^2\frac{\phi}{2}\Big].
\eeqa
\end{description}
 \textbf{\footnotesize T($k_4$)-P($\zeta_3,k_3$)-M($\epsilon_{\mu\nu},k_2$)-P($\zeta_1,k_1$)}\\
Since the result for $\cA_{TPMP}$ involves lengthy formula, we separate the equation into two parts.\\
The contribution from $\epsilon_{\mu\nu}$ $\Rightarrow$
\beqa
\notag\cA_{TPMP}&\sim&\frac{\Ga(-\al^\prime t-1)\Ga(-\al^\prime s-1)}{\Ga(-\al^\prime t-\al^\prime s+2)}\times\\
\notag&&\times
\left\{\begin{array}{r}
(\al^\prime s+1)(\al^\prime t+1)\left[\begin{array}{r}(\al^\prime t)(\al^\prime t-1)(-4\al^{\prime2}\zeta_1\cdot k_3 \zeta_3\cdot k_1\epsilon_{\mu\nu}k^\mu_1k^\nu_1)\\
+(\al^\prime t)(\al^\prime s)(8\al^{\prime2}\zeta_1\cdot k_3 \zeta_3\cdot k_1\epsilon_{\mu\nu}k_1^\mu k_3^\nu)\\
+(\al^\prime s)(\al^\prime s-1)(-4\al^{\prime2}\zeta_1\cdot k_3 \zeta_3\cdot k_1\epsilon_{\mu\nu}k^\mu_3k^\nu_3)\end{array}\right]\\
+(\al^\prime t+\al^\prime s-1)\left[\begin{array}{r}
(\al^\prime t+1)(\al^\prime t)(\al^\prime s+1)(-4\al^{\prime2}\zeta_1\cdot k_3\zeta_3\cdot k_2\epsilon_{\mu\nu}k^\mu_1k^\nu_1)\\
+(\al^\prime t+1)(\al^\prime s+1)(\al^\prime s)(
8\al^{\prime2}\zeta_1\cdot k_3 \zeta_3\cdot k_2\epsilon_{\mu\nu}k^\mu_1k^\nu_3)\\
+(\al^\prime s+1)(\al^\prime s)(\al^\prime s-1)(-4\al^{\prime2}\zeta_1\cdot k_3\zeta_3\cdot k_2\epsilon_{\mu\nu}k^\mu_3k^\nu_3)\end{array}\right]\end{array}\right\}\\
\notag&\sim&\frac{\Ga(-\al^\prime t-1)\Ga(-\al^\prime s-1)}{\Ga(-\al^\prime t-\al^\prime s+2)}(\al^\prime s)^2\times\\
\notag&&\times\Big[(\al^\prime t)^2(\epsilon_{\mu\nu}k^\mu_1k^\nu_1)-(\al^\prime t)(\al^\prime s)(
2\epsilon_{\mu\nu}k^\mu_1k^\nu_3)+(\al^\prime s)^2(\epsilon_{\mu\nu}k^\mu_3k^\nu_3)\Big](4\al^{\prime2}\zeta_1\cdot k_3 \zeta_3\cdot k_1).\\
\eeqa
\begin{description}
  \item[]$\al^L_{-1}\al^L_{-1}$:
  \beqa
  \notag\cA_{TPMP}(LL)&\sim&\frac{\Ga(-\al^\prime t-1)\Ga(-\al^\prime s-1)}{\Ga(-\al^\prime t-\al^\prime s+2)}(\al^\prime t+\al^\prime s)^2(\al^\prime s)^2(e^L_2\cdot k_1)^2(4\al^{\prime2}e^T_1\cdot k_3 e^T_3\cdot k_1)\\
  &\sim&\frac{\Ga(-\al^\prime t-1)\Ga(-\al^\prime s-1)}{\Ga(-\al^\prime t-\al^\prime s)}\Big[-2^{\frac{12}{2}}i\al^{\prime\frac{10}{2}}E^6V_0^4\sin^2\frac{\phi}{2}\cos^2\frac{\phi}{2}\Big].
  \eeqa
  \item[]$\al^L_{-1}\al^T_{-1}$:
  \beqa
  \notag\cA_{TPMP}(LT)&\sim&\frac{\Ga(-\al^\prime t-1)\Ga(-\al^\prime s-1)}{\Ga(-\al^\prime t-\al^\prime s+2)}(\al^\prime t+\al^\prime s)(\al^\prime s)^3(e^L_2\cdot k_1)(e^T_2\cdot k_3)(-4\al^{\prime2}e^T_1\cdot k_3 e^T_3\cdot k_1)\\
  &\sim&\frac{\Ga(-\al^\prime t-1)\Ga(-\al^\prime s-1)}{\Ga(-\al^\prime t-\al^\prime s)}
  \Big[2^{\frac{14}{2}}\al^{\prime\frac{9}{2}}E^6V_0^3\sin^3\frac{\phi}{2}\cos\frac{\phi}{2}\Big].
  \eeqa
  \item[]$\al^T_{-1}\al^T_{-1}$:
  \beqa
  \notag\cA_{TPMP}(TT)&\sim&\frac{\Ga(-\al^\prime t-1)\Ga(-\al^\prime s-1)}{\Ga(-\al^\prime t-\al^\prime s+2)}(\al^\prime s)^4(e^T_2\cdot k_3)^2(4\al^{\prime2}e^T_1\cdot k_3 e^T_3\cdot k_1)\\
  &\sim&\frac{\Ga(-\al^\prime t-1)\Ga(-\al^\prime s-1)}{\Ga(-\al^\prime t-\al^\prime s)}\bigg[2^{\frac{16}{2}}i\al^{\prime\frac{8}{2}}E^6V_0^2\sin^4\frac{\phi}{2}\bigg].
  \eeqa
\end{description}
The contribution from $\epsilon_{\mu}$ $\Rightarrow$
\beqa
\notag\cA_{TPMP}&\sim&\frac{\Ga(-\al^\prime t-1)\Ga(-\al^\prime s-1)}{\Ga(-\al^\prime t-\al^\prime s+2)}\times\\
\notag&&\times\left\{\begin{array}{r}(\al^\prime t+1)(\al^\prime s+1)
\left\{\begin{array}{r}(\al^\prime t)(\al^\prime t-1)\Big[(2\al^\prime)^{\frac{3}{2}}\zeta_1\cdot k_3 \zeta_3\cdot k_1\epsilon\cdot k_1\Big]\\
+(\al^\prime s)(\al^\prime s-1)\Big[(2\al^\prime)^{\frac{3}{2}}\zeta_1\cdot k_3 \zeta_3\cdot k_1\epsilon\cdot k_3\Big]\end{array}\right\}
\\
+(\al^\prime t+\al^\prime s-1)\left\{\begin{array}{r}(\al^\prime t+1)(\al^\prime t)(\al^\prime s+1)\Big[(2\al^\prime)^{\frac{3}{2}}\zeta_1\cdot k_3\zeta_3\cdot k_2\epsilon\cdot k_1\Big]\\
+(\al^\prime s+1)(\al^\prime s)(\al^\prime s-1)\Big[(2\al^\prime)^{\frac{3}{2}}\zeta_1\cdot k_3\zeta_3\cdot k_2\epsilon\cdot k_3
\Big]\end{array}\right\}\end{array}\right\}\\
\notag&\sim&
\frac{\Ga(-\al^\prime t-1)\Ga(-\al^\prime s-1)}{\Ga(-\al^\prime t-\al^\prime s)}\frac{(\al^\prime s)^2\Big[(\al^\prime t)^2(\epsilon\cdot k_1)+(\al^\prime s)^2(\epsilon\cdot k_3)\Big]\Big[-(2\al^\prime)^{\frac{3}{2}}\zeta_1\cdot k_3 \zeta_3\cdot k_1\Big]}{(-\al^\prime t-\al^\prime s)^2}.\\
\eeqa
\begin{description}
  \item[]$\al^L_{-2}$:
  \beqa
  \cA_{TPMP}(L)\sim\frac{\Ga(-\al^\prime t-1)\Ga(-\al^\prime s-1)}{\Ga(-\al^\prime t-\al^\prime s)}\Big[2^{\frac{9}{2}}\al^{\prime4}E^5V_0^3\big(1+\sin^2\frac{\phi}{2}\big)\sin\phi\tan\frac{\phi}{2}\Big].
  \eeqa
  \item[]$\al^T_{-2}$:
  \beqa
  \cA_{TPMP}(T)\sim\frac{\Ga(-\al^\prime t-1)\Ga(-\al^\prime s-1)}{\Ga(-\al^\prime t-\al^\prime s)}\Big[2^{\frac{13}{2}}i\al^{\prime\frac{7}{2}}E^5V_0^2\sin^2\frac{\phi}{2}\tan\frac{\phi}{2}\Big].
  \eeqa
\end{description} 

%% file: IV-C-4.tex
 \par
\begin{figure}[h]
\begin{center}\includegraphics[width=0.85\textwidth]{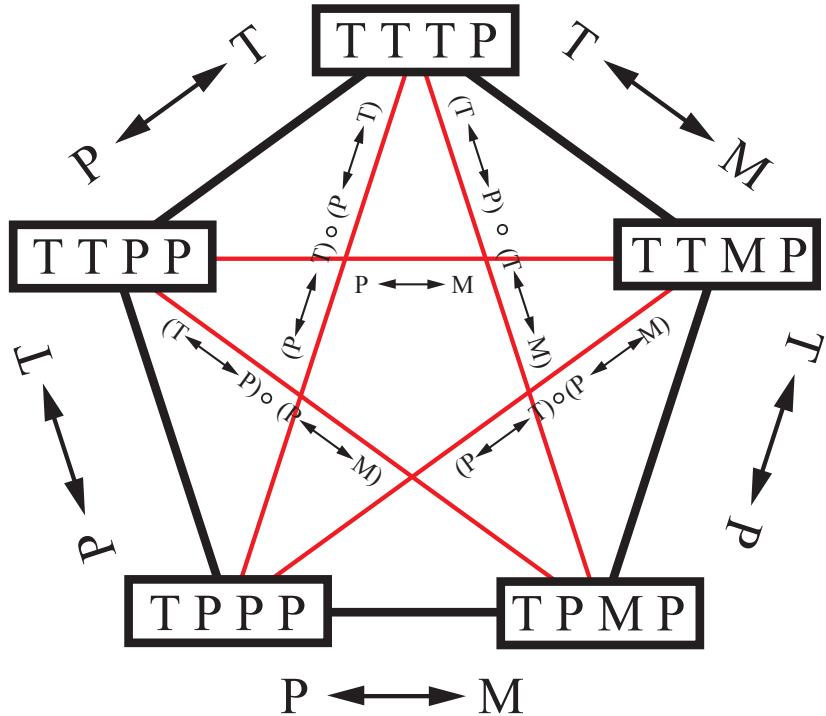}\end{center}
\caption{Replacement rules as a manifestation of inter-level stringy symmetry}\label{fig3}
\end{figure}

 From the five sample calculations of high-energy stringy scattering amplitudes in section IV-C-2, we observe three interesting features:
  \begin{enumerate}
    \item[(1)] The leading energy power $(\sqrt{\al^\prime}E)^n$ (e.g. $\cA_{TTMP}(LT)\sim(\sqrt{\al^\prime}E)^6$ is subleading) obeys the addition rule:
        \beqa
        n=\sum^{4}_{i=1}(N_i+1).
    \eeqa
    Here the level $N_i$ for each particle is: $N(T)=0$, $N(P)=1$, $N(M)=2$. This feature was first pointed out in \cite{Chan:2003ee,Chan:2004yz} and there is a saddle-point calculation \cite{Chan:2005ne,Chan:2005zp,Chan:2005ji} supporting that this addition rule is true for arbitrary high-energy stringy scattering amplitudes. It seems to suggest some partonic (or string-bit) picture of high-energy stringy scatterings.
    \item[(2)] Comparing different degrees of freedom at the same mass level, we see that there exists linear relations \cite{Chan:2003ee,Chan:2004yz} among high-energy stringy scattering amplitudes
        \beqa
        &&\cA_{TTMP}(LL):\cA_{TTMP}(TT)=1:4,\\
        &&\cA_{TPMP}(LL):\cA_{TPMP}(TT)=1:4.
        \eeqa
        In addition, there are also linear relations among high-energy stringy scattering amplitudes with the same total level $n$
        \beqa
        \cA_{TTMP}(LL):\cA_{TTMP}(TT):\cA_{TPPP}=1:4:-4.
        \eeqa
    \item[(3)] Our main concern in the present study is to explore the symmetry pattern for high-energy stringy scattering amplitudes of different total level. One of the early attempt for studying the subleading high-energy stringy scattering amplitudes can be found in \cite{Ho:2006zu}. Here we wish to push the idea further, and indeed we can identify a set of replacement rules (ignoring the phase factors) for relating different high-energy stringy scattering amplitudes. The replacement rule is summarized in Fig.\ref{fig3}. One can see clearly that from the figure, if we start from the amplitude $\cA_{TTTP}$, by replacing any particle (T or P) into a different particle, we can predict the results for new high-energy stringy scattering amplitudes based on the rules in Table \ref{5}.
  \end{enumerate}
  \par In contrast, as we move to the infinite light-like linear dilaton gradient limit $V_0/E\rightarrow\infty$, we find the three special features related to high-energy stringy symmetry are modified as follows:

  \begin{enumerate}
    \item [(1)] First of all, there is no reduction of energy power for the stringy scattering amplitudes involving $\al^L_{-1}$ oscillator. Consequently, the leading energy power $(\sqrt{\al^\prime}E)^n$ obeys different addition rule:
        \beqa
        n=\Big(\sum_{i=1}^4N_i\Big)+2
        \eeqa
        In addition, all high-energy stringy scattering amplitudes have a $(\sqrt{\al^\prime}V_0)^2$ factor if we take into account the state normalization Eqs.\eqref{PNSnomo} properly.
    \item [(2)] In the $V_0/E\rightarrow\infty$ limit, there is no linear relations among independent degrees of freedom at the same mass level. Indeed,
        \beqa
        &&\cA_{TTMP}(LL):\cA_{TTMP}(LT):\cA_{TTMP}(TT)=\cos^2\frac{\phi}{2}:i\sin\phi:-4\sin^2\frac{\phi}{2}\\
        &&\cA_{TPMP}(LL):\cA_{TPMP}(LT):\cA_{TPMP}(TT)=\cos^2\frac{\phi}{2}:i\sin\phi:-4\sin^2\frac{\phi}{2}.
        \eeqa
                This angle-dependent relation should not be understood as a lost of symmetry. Rather, if we examine the angle dependence of $\cA_{TTMP}(LT)$ and $\cA_{TPMP}(LT)$, we see that they are of subleading energy orders in the flat space-time, Eqs.\eqref{TTMPLT},\eqref{TPMPLT}, but now carry the same energy orders as those of $(LL)$ and $(TT)$ in the infinite dilation gradient case. In view of this, it would be natural to interpret that the flat space-time linear relations among high-energy stringy scattering amplitudes should be considered as degenerate limit of high-energy stringy symmetry. This is very similar to the analogy of group contraction we mentioned in the introduction. The isometry group for the Euclidean plane contains two commuting generators (translations along two coordinate axes), which are degenerate deformation of the (more symmetric, hence non-commuting) angular momentum generators.
    \item [(3)] Having explained the idea of deformation of the high-energy stringy symmetry, it should be clear that the separation of symmetry patterns among equal-mass states (a horizontal relation) and inter-level states (a vertical relation) in the flat space-time background is not really essential. In the simple example of bosonic open string theory in the light-like linear dilaton background, as we have studied in this paper, it is of more importance to see the general symmetry pattern as reflected in the replacement rules. By comparing the sample calculations of high-energy stringy scattering amplitudes, we can identify the replacement rules (ignoring the phase factors) at $V_0/E\rightarrow\infty$ in Table \ref{5}. While there are three identical rules regarding $T\rightarrow P$, $P\rightarrow M_{TT}$, and $T\rightarrow M_{TT}$ replacements, we find the deformation indeed modifies the rest of replacement rules at $V_0=0$ to a new set of replacement rules at $V_0/E\rightarrow\infty$. These new replacement rules at $V_0/E\rightarrow\infty$ fixed-point justify our proposal for the universal property of high-energy stringy symmetry.
  \end{enumerate}
\begin{table}[h]
\begin{center}
\begin{tabular}{|c|c|c|}
\hline
&~~~~~Flat Background~($V_0=0$, $E\rightarrow\infty$)~~~~&~Linear Dilaton Background~($V_0/E\rightarrow\infty$, $E\rightarrow\infty$)\\
\hline
$T\rightarrow P$&$\displaystyle2^{\frac{3}{2}}\sqrt{\al^\prime}E\tan\frac{\phi}{2}$&$\displaystyle2^{\frac{3}{2}}\sqrt{\al^\prime}E\tan\frac{\phi}{2}$\\
\hline
$P\rightarrow M_{LL}$&$\displaystyle2^{-\frac{1}{2}}\sqrt{\al^\prime}E\tan\frac{\phi}{2}$&$\displaystyle2^{-\frac{1}{2}}\al^{\prime\frac{3}{2}}EV_0^2\cot\frac{\phi}{2}$\\
\hline
$P\rightarrow M_{LT}$&$\displaystyle2^{-\frac{1}{2}}\frac{\cos\phi}{\cos^2\frac{\phi}{2}}$&$\displaystyle 2^{\frac{1}{2}}\al^\prime EV_0$\\
\hline
$P\rightarrow M_{TT}$&$\displaystyle2^{\frac{3}{2}}\sqrt{\al^\prime}E\tan\frac{\phi}{2}$&$\displaystyle2^{\frac{3}{2}}\sqrt{\al^\prime}E\tan\frac{\phi}{2}$\\
\hline
$P\rightarrow M_{L}$&0&$\displaystyle \frac{2i\sqrt{\al^\prime}V_0\big(1+\sin^2\frac{\phi}{2}\big)}{\sin\phi}$\\
\hline
$P\rightarrow M_{T}$&0&$\displaystyle2^{-\frac{1}{2}}\sec^2\frac{\phi}{2}$\\
\hline
$T\rightarrow M_{LL}$&$\displaystyle2\al^\prime E^2\tan^2\frac{\phi}{2}$&$2\al^{\prime2} E^2V_0^2$\\
\hline
$T\rightarrow M_{LT}$&$\displaystyle2\sqrt{\al^\prime} E\frac{\cos\phi}{\cos^2\frac{\phi}{2}}\tan\frac{\phi}{2}$&$\displaystyle2^2\al^{\prime\frac{3}{2}} E^2V_0\tan\frac{\phi}{2}$\\
\hline
$T\rightarrow M_{TT}$&$\displaystyle2^3\al^{\prime}E^2\tan^2\frac{\phi}{2}$&$\displaystyle2^3\al^{\prime}E^2\tan^2\frac{\phi}{2}$\\
\hline
$T\rightarrow M_{L}$&0&$\displaystyle \frac{2^{\frac{3}{2}}i\al^\prime EV_0\big(1+\sin^2\frac{\phi}{2}\big)}{\cos^2\frac{\phi}{2}}$\\
\hline
$T\rightarrow M_{T}$&0&$\displaystyle2\sqrt{\al^\prime}E\sec^2\frac{\phi}{2}\tan\frac{\phi}{2}$\\
\hline
\end{tabular}
\end{center}
\caption{Replacement rules for four-point functions}\label{5}
~\\ \end{table}

%% file: V.tex
\par In this paper, using bosonic open string theory in the light-like linear dilaton background as an illustration, we discuss the universal property of the high-energy stringy symmetry. This universal property is shown in two aspects:
\begin{enumerate}
  \item[(1)] An explicit formula for the covariant spectrum, up to the first massive level, is derived as a function of the moduli parameter, namely, the light-like linear dilaton gradient $V_0$.
  \item[(2)] The fixed-angle high-energy limits of various string scattering amplitudes are calculated as a function of the moduli parameter $V_0$.
\end{enumerate}
From these results, we identify two fixed points in the spectral flow and derive the replacement rules as a signature of deformed high-energy stringy symmetry. While this example is only one of many exactly solvable string theory models in which one can realize the idea of a universal stringy symmetry, one can follow the idea and explore new symmetry patterns in other conformal invariant backgrounds. It should be emphasized that our conclusion of the high-energy stringy
symmetry is based on the study of tree-level
stringy scattering amplitudes only. In principle, one can also
calculate the one-loop stringy scattering amplitudes
and extract their high-energy limits based on the operator formalism
\cite{Green:1987sp, Chan:2004ni}.
It would be of interest to see if the pattern of high-energy stringy
symmetry persists at the one-loop level as the original
claim \cite{Gross:1988ue}, or it can provide an independent check of the
results by Moeller and West \cite{Moeller:2005ez}.
\par In conclusion, we list some of the problems that we hope to finish and provide further insights toward an understanding of the nature of stringy symmetry:
\begin{enumerate}
  \item[(1)] Probing the deformation of the stringy symmetry in other kinematic regions (not necessary at high energies).
  \item[(2)] Identification of the symmetry generators based on a "tensionless" string approach to the properly scaled string world-sheet action \cite{Isberg:1993av,Sundborg:2000wp,Sezgin:2002rt,Bonelli:2003kh}.
  \item[(3)] Extending the formulae in Moore's work \cite{Moore:1993zc,Moore:1993qe,Moore:1994rm} on stringy symmetry and check the compatibility between inter-level Ward identities and the deformation of stringy spectrum via spectral flow.
  \item[(4)] Clarify the role of Liouville mode and extend our study on string symmetry to various space-time dimension.
  \item[(5)] Identifying the universal string symmetry in the M-theory context.
\end{enumerate}

%% file: AKG.tex
The authors wish to acknowledge Pei-Ming Ho, Jen-Chi Lee, Shunsuke Teraguchi, Yi Yang, and Chi-Hsien Yeh
 for early collaborations and many discussions in shaping the idea of this current work. C.T. is grateful to Pravina Borhade for collaboration on a long ongoing project on the high-energy stringy symmetry for compactified bosonic closed string theory, which antedates and stimulates the present work. We are also grateful to Pei-Ming Ho and Darren Shih for explaining some of the crucial points in their work. This work is supported by the National Science Council of Taiwan under the contract 96-2112-M-029-002-MY3 and the string focus group under National Center for Theoretical Sciences. 

%% file: HESSLLDB.bbl
\begin{thebibliography}{}
\bibitem{Gross:1987kza}
  D.~J.~Gross and P.~F.~Mende,
  ``The High-Energy Behavior of String Scattering Amplitudes,''
  Phys.\ Lett.\  B {\bf 197} (1987) 129.

\bibitem{Gross:1987ar}
  D.~J.~Gross and P.~F.~Mende,
  ``String Theory Beyond the Planck Scale,''
  Nucl.\ Phys.\  B {\bf 303} (1988) 407.

\bibitem{Gross:1989ge}
  D.~J.~Gross and J.~L.~Manes,
  ``The High-energy Behavior of Open String Scattering,''
  Nucl.\ Phys.\  B {\bf 326} (1989) 73.

\bibitem{Gross:1988ue}
  D.~J.~Gross,
  ``High-Energy Symmetries Of String Theory,''
  Phys.\ Rev.\ Lett.\  {\bf 60} (1988) 1229.


\bibitem{Moore:1993zc}
  G.~W.~Moore,
  ``Finite In All Directions,''
  arXiv:hep-th/9305139.


\bibitem{Moore:1993qe}
  G.~W.~Moore,
  ``Symmetries and symmetry breaking in string theory,''
  arXiv:hep-th/9308052.

\bibitem{Moore:1994rm}
  G.~W.~Moore,
  ``Addendum to: Symmetries of the bosonic string S matrix,''
  arXiv:hep-th/9404025.

\bibitem{Lee:2003vm}
  J.~C.~Lee,
  ``Stringy symmetries and their high-energy limit,''
  arXiv:hep-th/0303012.

\bibitem{Chan:2003ee}
  C.~T.~Chan and J.~C.~Lee,
  ``Stringy symmetries and their high-energy limits,''
  Phys.\ Lett.\  B {\bf 611} (2005) 193
  [arXiv:hep-th/0312226].

\bibitem{Chan:2004yz}
  C.~T.~Chan and J.~C.~Lee,
  ``Zero-norm states and high-energy symmetries of string theory,''
  Nucl.\ Phys.\  B {\bf 690} (2004) 3
  [arXiv:hep-th/0401133].

\bibitem{Chan:2004tb}
  C.~T.~Chan, P.~M.~Ho and J.~C.~Lee,
  ``Ward identities and high-energy scattering amplitudes in string theory,''
  Nucl.\ Phys.\  B {\bf 708} (2005) 99
  [arXiv:hep-th/0410194].

\bibitem{Chan:2005ne}
  C.~T.~Chan, P.~M.~Ho, J.~C.~Lee, S.~Teraguchi and Y.~Yang,
  ``Solving all 4-point correlation functions for bosonic open string  theory
  in the high energy limit,''
  Nucl.\ Phys.\  B {\bf 725} (2005) 352
  [arXiv:hep-th/0504138].

\bibitem{Chan:2005zp}
  C.~T.~Chan, P.~M.~Ho, J.~C.~Lee, S.~Teraguchi and Y.~Yang,
  ``High-energy zero-norm states and symmetries of string theory,''
  Phys.\ Rev.\ Lett.\  {\bf 96} (2006) 171601
  [arXiv:hep-th/0505035].

\bibitem{Chan:2005ji}
  C.~T.~Chan, P.~M.~Ho, J.~C.~Lee, S.~Teraguchi and Y.~Yang,
  ``Comments on the high energy limit of bosonic open string theory,''
  Nucl.\ Phys.\  B {\bf 749} (2006) 266
  [arXiv:hep-th/0509009].

\bibitem{Ho:2006zu}
  P.~M.~Ho and X.~Y.~Lin,
  ``Linear relations among 4-point functions in the high energy limit of string
  theory,''
  Phys.\ Rev.\  D {\bf 73} (2006) 126007
  [arXiv:hep-th/0604026].

\bibitem{Witten:1985cc}
  E.~Witten,
  ``Noncommutative Geometry And String Field Theory,''
  Nucl.\ Phys.\  B {\bf 268} (1986) 253.

\bibitem{Kao:2002us}
  H.~C.~Kao and J.~C.~Lee,
  ``Decoupling of degenerate positive-norm states in Witten's string field
  theory,''
  Phys.\ Rev.\  D {\bf 67} (2003) 086003
  [arXiv:hep-th/0212196].

\bibitem{Chan:2005qd}
  C.~T.~Chan, J.~C.~Lee and Y.~Yang,
  ``Anatomy of zero-norm states in string theory,''
  Phys.\ Rev.\  D {\bf 71} (2005) 086005
  [arXiv:hep-th/0501020].

\bibitem{P}
P.~M.~Ho, Private Communication. It is important to emphasize that in our calculations, the mass ratios among different stringy excitations is fixed. This distinguishes our approach from the standard tensionless string approach \cite{Isberg:1993av, Sundborg:2000wp, Sezgin:2002rt, Bonelli:2003kh}.

\bibitem{Cornwall:1974km}
  J.~M.~Cornwall, D.~N.~Levin and G.~Tiktopoulos,
  ``Derivation Of Gauge Invariance From High-Energy Unitarity Bounds On The S
  Matrix,''
  Phys.\ Rev.\  D {\bf 10} (1974) 1145
  [Erratum-ibid.\  D {\bf 11} (1975) 972].

\bibitem{Witten:1988sy}
  E.~Witten,
  ``The Search for Higher Symmetry in String Theory,''
  IASSNS-HEP-88/55.

\bibitem{Chan:2006qf}
  C.~T.~Chan, J.~C.~Lee and Y.~Yang,
  ``Scatterings of massive string states from D-brane and their linear
  relations at high energies,''
  Nucl.\ Phys.\  B {\bf 764} (2007) 1
  [arXiv:hep-th/0610062].

\bibitem{Chan:2006pf}
  C.~T.~Chan, J.~C.~Lee and Y.~Yang,
  ``Power-law Behavior of Strings Scattered from Domain-wall at High   Energies
  and Breakdown of their Linear Relations,''
  arXiv:hep-th/0610219.

\bibitem{Lee:2006fp}
  J.~C.~Lee and Y.~Yang,
  ``Linear relations of high energy absorption / emission amplitudes of
  D-brane,''
  Phys.\ Lett.\  B {\bf 646} (2007) 120
  [arXiv:hep-th/0612059].

\bibitem{Lee:2007cx}
  J.~C.~Lee and Y.~Yang,
  ``Linear Relations and their Breakdown in High Energy Massive String
  Scatterings in Compact Spaces,''
  Nucl.\ Phys.\  B {\bf 784} (2007) 22
  [arXiv:0705.1872 [hep-th]].

\bibitem{Lee:2007dp}
  J.~C.~Lee and Y.~Yang,
  ``High-energy Massive String Scatterings from Orientifold Planes,''
  Nucl.\ Phys.\  B {\bf 798} (2008) 198
  [arXiv:0712.4245 [hep-th]].

\bibitem{Lee:2008ba}
  J.~C.~Lee, T.~Takimi and Y.~Yang,
  ``High-energy String Scatterings of Compactified Open String,''
  Nucl.\ Phys.\  B {\bf 804} (2008) 250
  [arXiv:0805.3168 [hep-th]].

\bibitem{Ko:2008re}
  S.~L.~Ko, J.~C.~Lee and Y.~Yang,
  ``Kummer function and High energy String Scatterings,''
  arXiv:0811.4502 [hep-th].

\bibitem{Ko:2008ft}
  S.~L.~Ko, J.~C.~Lee and Y.~Yang,
  ``Pattens of High energy Massive String Scatterings in the Regge regime,''
  arXiv:0812.4190 [hep-th].

\bibitem{Ho:2007ar}
  P.~M.~Ho and S.~Y.~Shih,
  ``Discrete States in Light-Like Linear Dilaton Background,''
  JHEP {\bf 0801} (2008) 054
  [arXiv:0711.2792 [hep-th]].


\bibitem{Chan:2009xx}
  C.~T.~Chan and W.~M.~Chen,
  ``Vertex Operators and Scattering Amplitudes of the Bosonic Open String
 Theory in the Linear Dilaton Background,''
  arXiv:0907.5472 [hep-th].


\bibitem{Green:1987sp}
  M.~B.~Green, J.~H.~Schwarz and E.~Witten,
  ``Superstring Theory. Vol. 1: Introduction,''
{\it  Cambridge, Uk: Univ. Pr. ( 1987) 469 P. ( Cambridge Monographs On Mathematical Physics)}

\bibitem{Chan:2009vx}
  C.~T.~Chan and W.~M.~Chen,
  ``Massive Vertex Operators and Conformal Algebra of the Bosonic Open String
  Theory in Flat Space-time,''
  arXiv:0907.5347 [hep-th].

\bibitem{Lee:1989rc}
  J.~C.~Lee,
  ``New Symmetries of Higher Spin States in String Theory,''
  Phys.\ Lett.\  B {\bf 241} (1990) 336.

\bibitem{Lee:1994wp}
  J.~C.~Lee,
  ``Generalized On-Shell Ward Identities in String Theory,''
  Prog.\ Theor.\ Phys.\  {\bf 91} (1994) 353
  [arXiv:hep-th/0503005].

\bibitem{CCP}
P.~Borhade, C.~T.~Chan, W.~M.~Chen, work in progress.

\bibitem{Isberg:1993av}
  J.~Isberg, U.~Lindstrom, B.~Sundborg and G.~Theodoridis,
  ``Classical and quantized tensionless strings,''
  Nucl.\ Phys.\  B {\bf 411} (1994) 122
  [arXiv:hep-th/9307108].

\bibitem{Sundborg:2000wp}
  B.~Sundborg,
  ``Stringy gravity, interacting tensionless strings and massless higher
  spins,''
  Nucl.\ Phys.\ Proc.\ Suppl.\  {\bf 102} (2001) 113
  [arXiv:hep-th/0103247].

\bibitem{Sezgin:2002rt}
  E.~Sezgin and P.~Sundell,
  ``Massless higher spins and holography,''
  Nucl.\ Phys.\  B {\bf 644} (2002) 303
  [Erratum-ibid.\  B {\bf 660} (2003) 403]
  [arXiv:hep-th/0205131].

\bibitem{Bonelli:2003kh}
  G.~Bonelli,
  ``On the tensionless limit of bosonic strings, infinite symmetries and
  higher spins,''
  Nucl.\ Phys.\  B {\bf 669} (2003) 159
  [arXiv:hep-th/0305155].

\bibitem{Chan:2004ni}
  C.~T.~Chan and J.~C.~Lee,
  ``One-loop massive scattering amplitudes and Ward identities in string
  theory,''
  Prog.\ Theor.\ Phys.\  {\bf 115} (2006) 229
  [arXiv:hep-th/0411212].

\bibitem{Moeller:2005ez}
  N.~Moeller and P.~C.~West,
  ``Arbitrary four string scattering at high energy and fixed angle,''
  Nucl.\ Phys.\  B {\bf 729} (2005) 1
  [arXiv:hep-th/0507152].

\end{thebibliography}
